\documentclass[11pt]{article}

\usepackage{amsmath} % AMS Math Package
\usepackage{amsthm} % Theorem Formatting
\usepackage{amssymb}	% Math symbols such as \mathbb
\usepackage{graphicx} % Allows for eps images
\usepackage{multicol} % Allows for multiple columns
\usepackage{multirow}
\usepackage{color}
\usepackage[dvips,letterpaper,margin=1in,bottom=1in]{geometry}

\usepackage[utf8]{inputenc}
\usepackage[english]{babel}

\usepackage{diagbox}
\usepackage{mathtools}

\newtheorem{theorem}{Theorem}[section]

\newtheorem{lemma}[theorem]{Lemma}

\newtheorem{definition}{Definition}[section]
\newtheorem{example}{Example}
\newtheorem{remark}{Remark}[section]

\newcommand{\braket}[2]{\left< #1 \vphantom{#2} \middle| #2 \vphantom{#1} \right>} % for Dirac brackets

\DeclarePairedDelimiter\abs{\lvert}{\rvert}

\DeclarePairedDelimiter\ket{\lvert}{\rangle}
\DeclarePairedDelimiter\bra{\langle}{\rvert}

\usepackage{latexsym}
\usepackage{CJK}

\usepackage{enumerate}

\usepackage{algorithm}
\usepackage{algpseudocode}
\usepackage{stmaryrd}
\usepackage{booktabs}

\usepackage{hyperref}
\newcommand{\footremember}[2]{%
    \footnote{#2}
    \newcounter{#1}
    \setcounter{#1}{\value{footnote}}%
}

\usepackage{cite}
\usepackage{bbm}

\usepackage{tablefootnote}
\usepackage{threeparttable}
\usepackage{scrextend}

\usepackage{tikz}
\usetikzlibrary{positioning}

\usepackage{qcircuit}

\def\h{\ensuremath{\mathcal{H}}}

\begin{document}
	%\begin{CJK*}{GBK}{song}

    \title{Equivalence Checking of Sequential Quantum Circuits}
        \author{
            Qisheng Wang \footremember{1}{Qisheng Wang is with the Department of Computer Science and Technology, Tsinghua University, Beijing 100084, China (e-mail: \url{QishengWang1994@gmail.com}).}
            \and Riling Li \footremember{2}{Riling Li was with the Department of Computer Science and Technology, Tsinghua University, Beijing 100084, China. He is now with the State Key Laboratory of Computer Science, Institute of Software, Chinese Academy of Sciences, Beijing 100190, China (e-mail: \url{lirl@ios.ac.cn}).}
            \and Mingsheng Ying \footremember{3}{Mingsheng Ying is with the State Key Laboratory of Computer Science, Institute of Software, Chinese Academy of Sciences, Beijing 100190, China, and also with the Department of Computer Science and Technology, Tsinghua University, Beijing 100084, China (e-mail: \url{yingms@ios.ac.cn}).}
        }
        \date{}
        \maketitle

    \begin{abstract}
    We define a formal framework for equivalence checking of sequential quantum circuits. The model we adopt is a quantum state machine, which is a natural quantum generalisation of Mealy machines.
A major difficulty in checking quantum circuits (but not present in checking classical circuits) is that the state spaces of quantum circuits are continuums.
This difficulty is resolved by our main theorem showing that equivalence checking of two quantum Mealy machines can be done with input sequences that are taken from some chosen basis (which are finite) and have a length quadratic in the dimensions of the state Hilbert spaces of the machines. Based on this theoretical result, we develop an (\textit{and to the best of our knowledge, the first}) algorithm for checking equivalence of sequential quantum circuits with running time $\mathcal{O}(2^{3m+5l}(2^{3m}+2^{3l}))$, where $m$ and $l$ denote the numbers of input and internal qubits, respectively. The complexity of our algorithm is comparable with that of the known algorithms for checking classical sequential circuits in the sense that both are exponential in the number of (qu)bits. Several case studies and experiments are presented.
    \end{abstract}


    \newpage

    \tableofcontents
    \newpage

\section{Introduction}\label{sec:introduction}

\textbf{\textit{Hardware verification}} is emerging as an important issue in quantum computing with the recent rapid progress in quantum hardware implemented by industries like IBM, Google and Intel. A series of testing and verification techniques have already been developed for quantum circuits in the last fifteen years (see for example \cite{Ab06} \cite{Ber18} \cite{Hung04} \cite{Pal12} \cite{Pal18} \cite{Sei12} \cite{Wil09}).
More generally, Design Automation technology has been gradually introduced into quantum computing in the last few years (see for example \cite{Roet17} \cite{Martin17} \cite{Martin}).

\textbf{\textit{Equivalence checking}} is arguably the most important formal verification technique being employed in the design flows of classical computing hardware. Equivalence checking of quantum circuits was first studied in \cite{Via07} based on a quantum variant of BDDs (Binary Decision Diagrams), called QuIDD (Quantum Information Decision Diagram). The notion of miter circuit was generalised in \cite{Markov} to reversible miter so that various simplification techniques of quantum circuits can be used in equivalence checking. In \cite{Lu11}, a method for checking equivalence of multiple-valued quantum circuits was introduced by extending data structure XQDD (X-decomposition Quantum Decision Diagram) defined in \cite{Lu08}, In a series of papers \cite{Nie14} \cite{Bur20a} \cite{Bur20b}, a scheme for checking equivalence of quantum circuits has been systematically developed using data structure QMDD (Quantum Multiple-Valued Decision Diagram) \cite{Mil06} \cite{Nie17}. Another way for checking equivalence of quantum circuits was proposed in \cite{Amy18} employing (a finite variant of) Feynman path integral.

As is well-known, digital logic circuits are divided into two categories: \textbf{\textit{combinational circuits}} and \textbf{\textit{sequential circuits}}.
The output value of a combinational circuit is a function of only the current input value,
while
the output value of a sequential circuit depends on not only the external input value but also the stored internal information, which is designed to be fed back to the sequential circuit itself.
Almost all practical digital devices contain (classical) sequential circuits.

An early application area of the model of sequential quantum circuits is the quantum control theory \cite{WM09}, where a quantum system is controlled by feeding the information obtained from it back to itself. In the view of sequential logic, the quantum control behaves as a sequential machine that takes the last output state as input at each time step. In particular, it was shown in \cite{Llo00} \cite{NWCL00} that  coherent quantum feedback control that feeds quantum information coherently back to the system can perform some tasks that cannot be done by a controller processing classical information. Subsequently, the notion of quantum feedback in quantum control theory was extended to the model of quantum feedback networks \cite{GJ09a} \cite{GJ09b}, and a Quantum Hardware Description Language QHDL was designed in \cite{TNP+12} for specification of photonic quantum circuits with feedback.

Recently, sequential circuit models started to play an essential role in quantum computing; for example, sequential quantum circuits are employed in the RUS (Repeat-Until-Success) scheme \cite{Pae14} for efficient implementation of quantum logic gates.
\begin{example}\label{exam-rus} $V_3 = (I+2iZ)/\sqrt5$ is one of the six gates in the \textit{efficiently universal single-qubit gate set} \cite{Harrow02}. The RUS implementation of $V_3$ given in \cite{Pae14} is a sequential quantum circuit visualised in Fig. \ref{fig0} (the quantum gates used there will be explained in Section \ref{sec-comb}).
\begin{figure}[!htp]\centering
\normalsize\begin{equation*}\qquad \quad
\Qcircuit @C=.2em @R=.1em {
\lstick{q = \ket 0}    & \qw & \qw & \qw   & \qw & \gate{H} & \gate{T} & \gate{H} & \ctrl{1}     & \gate{T^\dag} & \gate{H} & \gate{T} & \ctrl{1}     & \gate{H} & \gate{T} & \gate{H} & \qw & \meter    & \cw  \\
\lstick{p = \ket \psi} &     & \qw & \qw > & \qw & \qw      & \qw      & \qw      & \gate{Z}     & \qw           & \qw      & \qw      & \gate{Z}     & \gate{Z} & \qw      & \qw      & \qw & \qw \\
              & \qwx  & \qw & \qw & \qw & \qw & \qw & \qw & \qw & \qw & \qw & \qw & \qw & \qw & \qw & \qw & \qw & \qw \qwx
              \gategroup{1}{6}{2}{16}{1.1em}{--}
}
\end{equation*}
\caption{A sequential quantum circuit for RUS implementation of $V_3$.}\label{fig0}
% \hrulefill \vspace*{4pt}
\end{figure}
At each step, it is supposed to implement $V_3$ with probability $5/8$ and identity with probability $3/8$. The circuit has two variables $q$ and $p$. The outcome of the measurement on $q$ is the output variable, indicating whether $V_3$ is successful implemented at the current step. Variable $p$ is used to store internal information.
In particular, this sequential circuit was \textit{physically implemented} on superconducting qubits~\cite{Rya17}.
\end{example}

\textbf{Contributions of This Paper}: The existing research on equivalence checking of quantum circuits focuses on their combinational logic. In this paper, we  define a formal framework for equivalence checking of sequential quantum circuits, where a natural quantum generalisation of Mealy machines is adopted to model sequential quantum circuits. Then the notion of equivalence of sequential quantum circuits can be formally defined in terms of the equivalence of quantum Mealy machines.

A major difference between quantum circuits and classical circuits is that the state spaces of quantum circuits are continuums, and in contrast the state spaces of classical circuits are finite.
In the case of combinational logic, it has been observed in the previous work mentioned above that equivalence checking of quantum circuits can be done with input sequences taken from some given basis (which are finite). To extend this idea to the case of sequential quantum logic, two additional dimensions of difficulty must be overcome: \begin{itemize}\item The dependence of the output of a sequential circuit on both the external input and the internal state complicates the procedure of reducing the equivalence checking to a specific basis of the state space; \item Quantum measurements in sequential circuits are essential when checking their equivalence; whereas the measurements in combinational circuits can always be pushed to the end and thus ignored in equivalence checking.\end{itemize} By circumventing these difficulties, we prove that equivalence checking of two quantum Mealy machines can be done with input sequences which are taken from a given basis and have a length quadratic in the dimensions of the state Hilbert spaces of the machines.
Based on it, we develop an algorithm for checking equivalence of two quantum Mealy machines (see Theorem \ref{thm-complexity}) with time complexity $\mathcal{O}(\abs{O}d_V d^5 (d^3 + d_\mathit{in}^3))$, where $\abs{O}$ is the number of possible measurement outcomes, $d_\mathit{in}$ and $d$ is the dimension of the input and internal Hilbert spaces, and $d_V$ is the dimension of the subspace of input (mixed) states that are being checked. When checking an arbitrary possible input, we take $d_V = \mathcal{O}(d_\mathit{in}^2)$. Through this, the equivalence checking of sequential quantum circuits with $m$ input qubits and $l$ internal qubits can be done in $\mathcal{O}(2^{3m+5l}(2^{3m}+2^{3l}))$ time. Our algorithm is then comparable with the known algorithms for equivalence checking of classical sequential circuits, which also has time complexity exponential in the number of bits.
%The complexity of our algorithm is comparable with that of the known algorithms for equivalence checking of classical sequential circuits.

To illustrate the effectiveness of our model and algorithm, we present several case studies and experiments, including testing of sequential quantum circuits in the above example and (i) quantum random walks, a tool widely used in quantum algorithm design (e.g. search and simulation); (ii) controlled quantum circuits (e.g. Toffoli gate); (iii) a circuit for testing quantum Fourier transforms; and (iv) quantum half adder. Although our examples only use several simple quantum gates, namely the Hadamard gate, CNOT gate, Toffoli gate and one-qubit rotations, the main results of this paper are not limited to these specific gates and can be applied to an arbitrarily given set of quantum gates.

\textbf{Related Work}: The work in \cite{Via07} \cite{Mil06} \cite{Markov} \cite{Lu11} \cite{Nie14} \cite{Amy18} mentioned in the above targets checking equivalence of general combinational quantum circuits. Another line of related research \cite{Gay13} \cite{Gay14} \cite{Gay18} but not mentioned above is concerned with equivalence checking of a special class of combinational quantum circuits used in quantum communication protocols, in particular those that can be expressed in the stabiliser formalism.

The formal model employed in this paper is quantum Mealy machines, a kind of quantum finite-state machines. There have been a large number of references for various models of quantum finite-state machines; two of the earliest papers in this area are \cite{Moo00} \cite{Kon97}; for recent surveys, see \cite{Amb15} \cite{Bha19}. The work presented in the following interesting papers are the closest to ours. Quantum finite-state machines were already used in \cite{Luk09} to represent quantum circuits, including sequential quantum circuits, although only several interesting examples are presented in \cite{Luk09}, without a formal description of quantum Mealy machines; in particular, equivalence checking of sequential quantum circuits was not considered there.
The models proposed in \cite{Li06} and \cite{Wang19} are similar to ours, but they are committed to accepting strings over a finite input alphabet. However, our model allows inputs to be strings over an infinite alphabet, i.e. sequences of quantum (mixed) states (although in a finite-dimensional Hilbert space). In fact, this extension of inputs to quantum finite-state machines is helpful for checking equivalence of sequential quantum circuits.

Technically, equivalence checking of sequential quantum circuits is resolved by reducing it to that of quantum Mealy machines. Equivalence checking of (weighted) automata has a long history. The first attempt on checking equivalence of probabilistic automata was made in \cite{Paz71}, with the observation that two probabilistic automata are equivalent if and only if they produce the same probability distribution on every input of length $\leq n_1 + n_2 - 1$, where $n_1$ and $n_2$ are the number of states in the two automata under checking, respectively.
Let us call the maximal length of inputs needed in equivalence checking \textit{the checking bound}. Then the above observation implies an equivalence checking algorithm with time complexity exponential in the number of states by directly checking every possible input of length within the checking  bound. A remarkable improvement was achieved in  \cite{Tze92} with  the first polynomial-time algorithm for equivalence checking of probabilistic automata.
Koshiba \cite{Kos01} and Liu and Qiu \cite{Li06} successfully extended the method of \cite{Tze92} to quantum automata.
They developed efficient algorithms for equivalence checking of various types of quantum automata with time complexity $O(n^8)$ with checking bound $(n_1+n_2)^2$, where $n_1,n_2$ are the respective dimensions of the two quantum automata under checking, and
$n = n_1 + n_2$.
This checking bound was later improved in \cite{LQ09} to $n_1^2+n_2^2-1$. Recently, the techniques for  equivalence checking of quantum automata were further extended in \cite{Wang19} to the case of mixed quantum states, and the time complexity was improved from $O(n^8)$ to $O(n^6)$. The main idea of our method for checking equivalence of quantum Mealy machines
% presented
in this paper is based on the  techniques mentioned above, especially those in \cite{Li06} and \cite{Wang19}, where the machine  models have a similar structure to ours.

\textbf{Organisation of the Paper}: For convenience of the reader, we briefly recall basics of combinational quantum circuits in Section \ref{sec-comb}. A formal description of sequential quantum circuits is given in Section \ref{sec-seq}.
%Several examples are presented there to show that quantum random walks, controlled quantum circuits and quantum Fourier transforms can be conveniently implemented as sequential quantum circuits.
The notion of quantum Mealy machine is introduced in Section \ref{sec-Mealy} as an abstract model of sequential quantum circuits.
%The applicability of this notion is illustrated by modelling the examples given in Sec. \ref{sec-seq}.
An algorithm for equivalence checking of quantum Mealy machines is developed in Section \ref{sec-checking}. Case studies are given in Section \ref{sec-case}. The experiments of executing the algorithm on the examples mentioned are briefly discussed in Section \ref{sec-experiment}. A brief conclusion is drawn in Section \ref{sec-con}.

\section{Combinational Quantum Circuits}\label{sec-comb}

In this section, we briefly review the basics of combinational quantum circuits; for more details, we refer to \cite{NC00}.

\subsection{Qubits}
The basic unit of quantum information is called a quantum bit or qubit. For each qubit $q$, we write $\mathcal{H}_{q}$ for its state Hilbert space, which is two-dimensional. Using the Dirac notation, a (pure) state of $q$ is represented by $|\psi\rangle=\alpha_0|0\rangle+\alpha_1|1\rangle$ with complex numbers $\alpha_0$ and $\alpha_1$ satisfying the normalisation condition $|\alpha_0|^2+|\alpha_1|^2=1;$ for example, $q$ can be in not only the basis states $|0\rangle, |1\rangle$ but also a superposition of them like: $|+\rangle=\frac{1}{\sqrt{2}}(|0\rangle+|1\rangle),\ |-\rangle=\frac{1}{\sqrt{2}}(|0\rangle-|1\rangle).$

A sequence $\overline{q}=q_1,...,q_n$ of distinct qubit variables is called a quantum register. Its state Hilbert space is the tensor product $\mathcal{H}_{\overline{q}}=\bigotimes_{i=1}^n\mathcal{H}_{q_i},$ which is $2^n$-dimensional. To explicitly describe a state of quantum register $\overline{q}$, let an integer $0\leq x<2^n$ be represented by a string $x_1...x_{n}\in\{0,1\}^n$ of $n$ bits: $x=\sum_{i=1}^{n}x_{i}\cdot 2^{n-i}.$ We shall not distinguish the integer $x$ from its binary representation. Thus, each (pure) state in $\mathcal{H}_{\overline{q}}$ can be written as:
$|\psi\rangle=\sum_{x=0}^{2^n-1}\alpha_x |x\rangle,$ where $\{|x\rangle\}$ is called the computational basis, and complex numbers $\alpha_x$ satisfy the normalisation condition $\sum_{x=0}^{2^n-1}|\alpha_x|^2=1.$ This state can also be represented by the $2^n$-dimensional column vector $|\psi\rangle=(\alpha_0,...,\alpha_{2^n-1})^T,$ where $T$ stands for transpose. We write $\bra\psi = (\alpha_0^*,...,\alpha_{2^n-1}^*)$ to denote the conjugate transpose of $\ket\psi$, and $\braket\psi\phi$ to denote the inner product of $\ket{\psi}$ and $\ket\phi$. For example, two qubits $q_1,q_2$ can be in an entangled state like the EPR (Einstein-Podolsky-Rosen) pair: $|\beta\rangle=\frac{1}{\sqrt{2}}(|00\rangle+|11\rangle).$

\subsection{Unitary Transformations}

A transformation of a quantum register $\overline{q}=q_1,...,q_n$ is modelled by a $2^n\times 2^n$ unitary matrix $U=\left(u_{ij}\right)$ satisfying $U^\dag U=I$, where $U^\dag$ stands for the conjugate transpose of $U$ and $I$ the unit matrix. Thus, state $|\psi\rangle$ is transformed by $U$ to a new state represented by column vector $|\psi^\prime\rangle=U|\psi\rangle$ (matrix multiplication), i.e. $|\psi^\prime\rangle=(\alpha_0^\prime,...,\alpha_{2^n-1}^\prime)^T$ where $\alpha_i^\prime=\sum_{j=0}^{2^n-1}u_{ij}\alpha_j$ for each $i=0,...,2^n-1$.
\begin{example}Some unitary transformations used in quantum computing: \begin{itemize}
\item Single-qubit transformations: Hadamard matrix $H=\frac{1}{\sqrt{2}}\left(\begin{array}{cc}1 & 1\\ 1 & -1\end{array}\right)$, $\pi/8$ gate $T=\left(\begin{array}{cc}1 & 0\\ 0 & e^{i\pi/4}\end{array}\right)$, and Pauli matrices $X=\left(\begin{array}{cc}0 & 1\\ 1 & 0\end{array}\right)$, $Y=\left(\begin{array}{cc}0 & -i\\ i & 0\end{array}\right)$, $Z=\left(\begin{array}{cc}1 & 0\\ 0 & -1\end{array}\right)$.
\iffalse
\begin{align*}
&{\rm Hadamard\ matrix}\
H=\frac{1}{\sqrt{2}}\left(\begin{array}{cc}1 & 1\\ 1 & -1\end{array}\right); \\
& \frac{\pi}{8}\ {\rm gate}\ T=\left(\begin{array}{cc}1 & 0\\ 0 & e^{i\pi/4}\end{array}\right);\\ &{\rm Pauli\ matrices}\ I=\left(\begin{array}{cc}1 & 0\\ 0 & 1\end{array}\right),\quad X=\left(\begin{array}{cc}0 & 1\\ 1 & 0\end{array}\right), \\
& Y=\left(\begin{array}{cc}0 & -i\\ i & 0\end{array}\right),\quad Z=\left(\begin{array}{cc}1 & 0\\ 0 & -1\end{array}\right).\end{align*}
\fi
For instance, $H$ transforms the basis states $|0\rangle, |1\rangle$ to $|+\rangle, |-\rangle$, respectively; that is, $H|0\rangle=|+\rangle$ and $H|1\rangle=|-\rangle$. Similarly, $Y|1\rangle=-i|0\rangle$, $Z|1\rangle=-|1\rangle$.
\item The $4\times 4$ matrix $\mathit{CNOT} = \left(\begin{array}{cc}I & 0\\ 0 & X\end{array}\right)$ represents a unitary transformation of two qubits, called the controlled NOT. It can generate entanglement; for example, it transforms separable state $|+\rangle|0\rangle$ to the EPR pair $|\beta\rangle$.\end{itemize}
\end{example}

\subsection{Quantum Measurements}
According to a basic postulate of quantum mechanics, a way of reading out data from a quantum system is a quantum measurement. In quantum circuits, we mainly use the measurements in the computational basis.
Let $\overline{q}=q_1,...,q_n$ be a quantum register. It is divided into two segments $\overline{p}=q_1,...,q_m$ $(m\leq n)$ and $\overline{r}=q_{m+1},...,q_n$. If $\overline{q}$ is in state $|\psi\rangle=\sum_{x\in\{0,1\}^n}\alpha_x|x\rangle$ and we perform a measurement on the first segment $\overline{p}$ in the computational basis, then output $a=(a_1,...,a_m)\in\{0,1\}^m$ is obtained with probability
$p(a)=\sum_{x\ {\rm s.t.}\ x_i=a_i\ (1\leq i\leq m)}|\alpha_x|^2$
and after that, the state of $\overline{r}$ becomes
$|\varphi_a\rangle=\frac{1}{\sqrt{p(a)}}\sum_{y\in\{0,1\}^{n-m}}\alpha_{ay}|y\rangle$,
where $ay\in\{0,1\}^n$ is the concatenation of $a$ and $y$. For example, if two qubits $q_1,q_2$ are in the entangled state $|\beta\rangle$, and we measure $q_1$ in the computational basis, then we obtain outcome $0$ with probability $\frac{1}{2}$ and after that, $q_2$ is in the basis state $|0\rangle$, and we obtain outcome $1$ also with probability $\frac{1}{2}$ and after that, $q_2$ is in state $|1\rangle$.

\subsection{Quantum Gates and Quantum Circuits}
Now we are ready to formally define the notions of quantum gates and quantum circuits.
Here, we adopt the algebraic language defined in \cite{YF10} \cite{Ying16}.

\begin{definition}[Quantum gates] For any positive integer $n$, if $U$ is a $2^{n}\times 2^{n}$
unitary matrix, and $q_1,...,q_n$ are qubit variables, then $G\equiv U[q_1,...,q_n]$
is called an $n$-qubit gate.
\end{definition}

\begin{example}[Hadamard gate]\label{ex-single}
 Whenever the Hadamard matrix $H$ acts on a qubit $q$, then we have the Hadamard gate $H[q]$.
If $q$ is currently in state $|1\rangle$, it will be in state $|-\rangle$ at the end of the Hadamard gate.
\end{example}

\begin{example}[Classically controlled quantum gates]
\begin{enumerate}\item Let $q_1,q_2$ be qubit variables. Then CNOT gate $\mathit{CNOT}[q_1,q_2]$ is a two-qubit gate with $q_1$ as the control qubit and $q_2$ as the target qubit. It acts as follows:
$\mathit{CNOT}[q_1,q_2]|i_1, i_2\rangle =|i_1, i_1\oplus i_2\rangle$
for $i_1,i_2\in\{0,1\},$ where $\oplus$ is addition modulo $2$; that is, if $q_1$ is set to $|1\rangle$, then $q_2$ is flipped, otherwise $q_2$ is left unchanged.

  \item Let $q_1, q_2, q_3$ be qubit variables. Then the Toffoli gate is defined by
  $
    \mathit{Toffoli}[q_1, q_2, q_3]\ket{i_1, i_2, i_3} = \ket{i_1, i_2, i_3 \oplus i_1i_2}
  $
  for $i_1, i_2, i_3 \in \{0, 1\}$; that is, $q_3$ is flipped if and only if $q_1$ and $q_2$ are both set to $|1\rangle$.

  \item More generally, suppose that $c_1, \dots, c_k \in \{0,1\}$, $U$ is a $2^n \times 2^n$ unitary matrix, and $p_1, \dots, p_k$ and $q_1, \dots, q_n$ are qubit variables. Then we can define a controlled gate:
  \[
        G \equiv \mathrm{C}^{c_1} \mathrm{C}^{c_2} \dots \mathrm{C}^{c_k} U [p_1, p_2, \dots, p_k, q_1, q_2, \dots, q_n]
  \]
    by
  \begin{align*}
      G \ket{i_1, \dots, i_k, j_1, \dots, j_n}=
      \begin{cases} \ket{i_1, \dots, i_k} U\ket{j_1, \dots, j_n}
    &{\rm if}\ i_l = c_l\ {\rm for}\ 1 \leq l \leq k;\\ \ket{i_1, \dots, i_k, j_1, \dots, j_n} &{\rm otherwise},\end{cases}
  \end{align*}
 where the first $k$ variables $p_1,...,p_k$ are control qubits, and the last $n$ variables $q_1,...,q_n$ are target qubits. In particular, we have: $
    \mathit{CNOT}[q_1, q_2] = \mathrm{C}^1 X [q_1, q_2],
  \mathit{Toffoli}[q_1, q_2, q_3]
     = \mathrm{C}^{1} \mathrm{C}^{1} X [q_1, q_2, q_3].
  $
\end{enumerate}
\end{example}

\begin{definition}[Quantum circuits]\label{circuit} A quantum circuit is a sequence of quantum gates: $C\equiv G_1...G_d,$ where $d\geq 1$ and $G_1,...,G_d$ are quantum gates.\end{definition}

For a quantum circuit $C=G_1...G_d$, let $\overline{q}=q_1,...,q_n$ be all of the qubit variables occurring in $C$. Then for each $1\leq i\leq d$, $G_i$ is modelled by a unitary matrix $U_i$ acting on a substring of $\overline{q}$. We often consider $U_i$ as a transformation on the whole string $\overline{q}$; that is, $U_i$ is identified with its cylindrical extension in $\mathcal{H}_{\overline{q}}$. Therefore, $C$ is represented by unitary matrix $U=U_d...U_1$.

%Now we can define the notion of equivalence between quantum circuits.

\begin{definition}[Equivalence of Circuits]\label{circuit-equivalence}Let $C_1,C_2$ be quantum circuits and $\overline{q}$ be the sequence of qubit variables occurring in $C_1$ and $C_2$. Then
$C_1$ and $C_2$ are called equivalent, written $C_1=C_2$, if we have: $C_1|\psi\rangle=C_2|\psi\rangle$ for any $|\psi\rangle\in\mathcal{H}_{\overline{q}}$.\end{definition}

\begin{example} The $\mathit{SWAP}$ gate is defined by $\mathit{SWAP}[q_1,q_2]|i_1,i_2\rangle=|i_2,i_1\rangle$ for $i_1,i_2\in\{0,1\}$, which swaps the states of two qubits $q_1,q_2$.
It can be implemented by the following circuit consisting of three CNOT gates:
$\mathit{SWAP}[q_1,q_2]=\mathit{CNOT}[q_1,q_2]\mathit{CNOT}[q_2,q_1]\mathit{CNOT}[q_1,q_2].$
\end{example}

%As mentioned in Section \ref{sec:introduction}, several algorithms have been proposed in the previous literature for checking equivalence of combinational quantum circuits.

\section{Sequential Quantum Circuits}\label{sec-seq}

In this section, we extend the model of combinational quantum circuits defined in the last section to sequential quantum circuits.
%\riling{'This paragraph needs further rivision as we add a subsection of physical realization'.}
%\delete{To demonstrate its applicability, we show how it can be used to describe quantum random walks, controlled quantum circuits and quantum Fourier transforms.}
% In particular, the feasibility of our sequential model on quantum hardware is briefly discussed in Sec. \ref{sec:feasibility}.

\subsection{Definition and Examples}\label{defSQC}

Sequential circuits can be further divided into two subcategories: synchronous logic and asynchronous logic. As the first step, we only consider synchronous sequential quantum circuits in this paper.
Recall from \cite{Jha} that the output values of a sequential (classical) circuit depend both on the current inputs and on stored internal information (and thus also on the past inputs). So, a sequential circuit is a combinational circuit with memory. It is capable of storing data and also performing certain logical operations upon this data. In other words, the output values of a sequential circuit at a given time is then a function of external input values and the stored data at that time. Such a circuit model has a natural quantum generalisation:

\begin{definition}A (synchronous) sequential quantum circuit $S$ consists of:\begin{itemize}\item $m$ input variables $q_1,...,q_m$, each of which is a qubit;
\item $m$ output variables $x_1,...,x_m$, each of which is a classical binary variable, taking value $0$ or $1$;
\item $l$ state variables $p_{1},...,p_l$, each of which is also a qubit;
\item a combinational quantum circuit $C$ with $n=m+l$ qubits $q_1,...,q_m,p_1,...,p_l$;
%\item $l$ memory elements, the $i$th of which is used to store the (quantum) state of variable $p_i$. \riling{? Is this item necessary? It says the same thing as the second item 'l state variables'. I think it can be absorbed into the 2rd item of this definition.}
\end{itemize}\end{definition}

A sequential quantum circuit is visualised in Fig. \ref{fig1}.
\begin{figure}[!htp]\centering
\normalsize\begin{equation*}
\Qcircuit @C=0.7em @R=.5em {
\lstick{q_1} & \qw & \qw & \multigate{5}{\text{A Combinational Quantum Circuit}} & \meter & \rstick{x_1} \cw \\
\lstick{\vdots} & \qw & \qw & \ghost{\text{A Combinational Quantum Circuit}} & \meter & \rstick{\vdots} \cw \\
\lstick{q_m} & \qw & \qw & \ghost{\text{A Combinational Quantum Circuit}} & \meter & \rstick{x_m} \cw \\
\lstick{p_1} \qwx[4] & \qw & \qw > & \ghost{\text{A Combinational Quantum Circuit}} & \qw & \qw \qwx[4] \\
\lstick{\vdots} & \\
\lstick{p_l} & \qwx[1] & \qw > & \ghost{\text{A Combinational Quantum Circuit}} & \qw \qwx[1] & & \\
&  & \qw & \qw & \qw & & & \\
& \qw & \qw & \qw & \qw & \qw & & \\
}
\end{equation*}
\caption{A synchronous sequential quantum circuit.}\label{fig1}
%\hrulefill \vspace*{4pt}
\end{figure}
Intuitively, as in a classical synchronous circuit, a clock produces a sequence of clock signals. Each signal is distributed to all of the memory elements. The output of each memory element only changes when triggered by the clock signal. The quantum hardware implementing a sequential circuit works very differently from its classical counterpart (\textit{See Appendix \ref{sec:feasibility} for a discussion about physical implementation of the ``memory'' and ``clock'' in a sequential quantum circuit}). Another difference between classical and quantum models is that in the quantum case, a measurement appears at the end of each qubit $q_i$, which is needed to produce a (classical) outcome $x_i$. To show reasonableness of the above model, let us consider two examples:

\begin{example}\label{exam-rus+1} The RUS (Repeat-Until-Success) implementation of quantum gate $V_3 = (I+2iZ)/\sqrt5$ given in Example \ref{exam-rus} can be formally described as a sequential quantum circuits with the following combinational part:
\begin{equation}\label{eq-rus}
\begin{aligned}
\mathcal{C}_\mathit{rus} =
     & H[q] T[q] H[q] \mathrm{C}^1Z[q, p]
     T^\dag[q] H[q] T[q] \\
     & \mathrm{C}^1Z[q, p]
     H[q] Z[p] T[q] H[q].
\end{aligned}
\end{equation}
\end{example}

\begin{example}\label{exam-walk} Quantum random walks are quantum analogues of random walks, with many applications in designing quantum algorithms. Here, we consider a quantum particle moving on a $4$-circle \cite{Kem09}:
\begin{enumerate}\item The circle has 4 different positions: $0, 1, 2, 3$, and thus the position space is a $4$-dimensional Hilbert space $\mathcal{H}_{p} = \operatorname{span} \{ \ket {0}_p, \ket {1}_p, \ket {2}_p, \ket {3}_p \}$, that is, the complex vector space with $\{ \ket 0, \ket 1, \ket 2, \ket 3 \}$ as an orthonormal basis.
In contrast to the classical case, the particle can be in a superposition of positions $0,1,2,3$, written $\alpha_0|0\rangle+\alpha_1|1\rangle+\alpha_2|2\rangle+\alpha_3|3\rangle$.
\item A quantum coin is used to control the walking direction. Its state space is $\mathcal{H}_c=\operatorname{span}\{\ket {0}_c,\ket {1}_c\}$, where $0, 1$ denote direction \textit{right} and \textit{left}, respectively. In contrast to the classical case, the coin can be in a superposition $\beta_0|0\rangle+\beta_1|1\rangle$, meaning the particle moving to left and right simultaneously.
\item In each step of the walk, the coin is flipped with a unitary operator $C$, and then the conditional translation of position is performed: $T\ket {0}_c \ket{i}_p=\ket {0}_c\ket{i+1}_p$ and $T\ket {1}_c \ket{i}_p=\ket {1}_c\ket{ i-1}_p$ for $i=0,1,2,3$, where $i+1$ and $i-1$ are computed modulo $4$.
Here, we use the subscripts $c, p$ to indicate the states of coin and position, respectively. Intuitively, if the coin is in state $0$ (resp. $1$), then the position is changed from $i$ to $i+1$ (resp. $i-1$).
%Equivalently, translation operator $S$ can be written as:
%\begin{align*}
%    S  = &\ket {0}_c \bra {0} \otimes \sum_i \ket{i+1}_p \bra {i}
%  + \ket {1}_c \bra {1} \otimes \sum_i \ket{i-1}_p\bra {i};
%\end{align*}
\item A qubit $d$ together with a measurement is set to detect whether position $3$ is reached.
%The detective qubit $d$ is measured in the computational basis.
Whenever position $3$ is reached, the detective qubit $d$ is flipped.
\end{enumerate}
As usual, we can employ two qubits $p_1$ and $p_2$ to encode the $4$ positions: $\ket{0}_p=\ket{0}_{p_1}\ket{0}_{p_2}, \ket {1}_p=\ket{ 0}_{p_1}\ket{ 1}_{p_2}, \ket {2}_p=\ket{ 1}_{p_1}\ket{ 0}_{p_2}, \ket {3}_p=\ket{ 1}_{p_1}\ket{ 1}_{p_2}$.
\begin{figure}[!htp]\centering
\normalsize\begin{equation*}
\Qcircuit @C=0.7em @R=.5em {
\lstick{d}            & \qw & \qw & \qw & \qw & \qw         & \qw               & \targ     & \qw & \meter    & \cw  \\
\lstick{c}        &     & \qw & \qw & \qw > & \gate{C}    & \multigate{2}{T}  & \qw       & \qw       & \qw & \qw  \\
\lstick{p_1}  & \qwx&     & \qw & \qw > & \qw         & \ghost{T}         & \ctrl{-2} & \qw       & \qw & \qwx\\
\lstick{p_2}  & \qwx& \qwx&     & \qw > & \qw         & \ghost{T}         & \ctrl{-3} & \qw       & \qwx& \qwx \\
                            & \qwx& \qwx& \qwx& \qw & \qw         & \qw               & \qw       & \qw \qwx  & \qwx&\qwx \\
                            & \qwx& \qwx& \qw & \qw & \qw         & \qw               & \qw       & \qw       & \qw \qwx&\qwx \\
                            & \qwx& \qw & \qw & \qw & \qw         & \qw               & \qw       & \qw       & \qw & \qw \qwx
                            \gategroup{1}{6}{4}{8}{1.8em}{--}
}
\end{equation*}
\caption{A sequential circuit for detecting quantum random walk.}\label{fig2}
% \hrulefill \vspace*{4pt}
\end{figure}
Then the walk can be implemented as a sequential quantum circuit shown in Fig. \ref{fig2}. This circuit has four variables $d, c, p_1$ and $p_2$, where $d$ is the only input variable, and $c, p_1$ and $p_2$ are state variables. The outcome of measurement on $d$ is the output variable. The combinational part of the circuit is defined by
\begin{equation} \label{eq-walk}
    \mathcal{C}_C = C[c] T[c,p_1,p_2] \mathit{Toffoli}[p_1, p_2, d].
\end{equation}
\end{example}

\subsection{Computation Process}
Now we give a formal description of the computational process of a sequential quantum circuit. From the discussion in the last section, the combinational circuit $C$ with $n$ qubits is represented by a $2^n\times 2^n$ unitary matrix $U=\left(U_{uv}\right)_{u,v=0}^{2^n-1}.$ Moreover, we need the following notations: for any column vector $|\psi\rangle=(\alpha_0, \alpha_1,...,\alpha_{2^n-1})^T$, its conjugate transpose is row vector $\langle\psi|=(\alpha^\ast_0,\alpha_1^\ast,...,\alpha_{2^n-1}^\ast)$. Thus, the inner product of two vectors $|\psi\rangle, |\varphi\rangle$ is $\langle\psi|\varphi\rangle$ (as an ordinary matrix multiplication), and matrix entries $U_{uv} =\langle u|U|v\rangle$ if we identify integers $u,v\in\left\{0,1,...,2^n-1\right\}$ with their binary representations $u,v\in\{0,1\}^n$.

Let $|\varphi_1\rangle,..., |\varphi_i\rangle, ...$ in $\mathcal{H}_2^{\otimes m}$ (the tensor product of $m$ copies of the $2$-dimensional Hilbert space) be an input sequence and $|\psi_0\rangle$ in $\mathcal{H}_2^{\otimes l}$ an initial state. Then the sequential circuit $S$ behaves as follows:

\iffalse
\begin{itemize}
\item \textbf{Step} 1: Input state $$|\varphi_1\rangle=\sum_{x\in\{0,1\}^m}\alpha_{1x}|x\rangle\in \mathcal{H}_2^{\otimes m}$$ to variables $q_1,...,q_m$. Variables $p_1,...,p_l$ are initialised to state $$|\psi_0\rangle=\sum_{y\in\{0,1\}^l}\beta_{0y}|y\rangle\in\mathcal{H}_2^{\otimes l}.$$ Applying unitary $U$ on $q_1,...,q_m,p_1,...,p_l$ yields:
\begin{align*}
    & U|\varphi_1\rangle|\psi_0\rangle = \sum_{a\in\{0,1\}^m,b\in\{0,1\}^l} \\
    & \qquad \left(\sum_{x\in\{0,1\}^m,y\in\{0,1\}^l}\alpha_{1x}\beta_{0y}\langle ab|U|xy\rangle\right)|ab\rangle.
\end{align*}
Measure $q_1,...,q_m$ in the computational basis. Then a measurement outcome $a_1=a_{11},...,a_{1m}\in\{0,1\}^m$ is output through variables $x_1,...,x_m$, respectively, with probability
\begin{equation}\label{c-prob1}
\begin{aligned}
& p(a_1)= \|\langle a_1|U|\varphi_1\rangle|\psi_0\rangle\|^2
= \\
& \sum_{b\in\{0,1\}^l} \left| \sum_{x\in\{0,1\}^m,y\in\{0,1\}^l}\alpha_{1x}\beta_{0y}\langle a_1b|U|xy\rangle\right|^2.
\end{aligned}
\end{equation}
After that, variable $p_1,...,p_l$ is in state
\begin{equation}\label{c-state1}
\begin{aligned}
    |\psi_1\rangle & =\frac{\langle a_1|U|\varphi_1\rangle|\psi_0\rangle}{\sqrt{p(a_1)}}
=\frac{1}{\sqrt{p(a_1)}}\sum_{b\in\{0,1\}^l} \\
    &\left(\sum_{x\in\{0,1\}^m,y\in\{0,1\}^l}\alpha_{1x}\beta_{0y}\langle a_1b|U|xy\rangle\right)|b\rangle.
\end{aligned}
\end{equation}
This state is then input to the memory elements and stored there.
\end{itemize}
\fi

\textbf{Step} $i$ $(i\geq 1)$: Input state
$|\varphi_i\rangle=\sum_{x\in\{0,1\}^m}\alpha_{ix}|x\rangle\in \mathcal{H}_2^{\otimes m}$
to variables $q_1,...,q_m$. Note that at the end of step $i-1$, variables $p_1,...,p_l$ are in state $|\psi_{i-1}\rangle\in\mathcal{H}_2^{\otimes l}$, and this state is stored in the memory elements. Now it is output from the memory elements to the combinational circuit. Assume that
$|\psi_{i-1}\rangle=\sum_{y\in\{0,1\}^l}\beta_{(i-1)y}|y\rangle.$
Unitary $U$ transforms the state of $q_1,...,q_m,p_1,...,p_l$ to:
\begin{align*}
    U|\varphi_i\rangle|\psi_{i-1}\rangle=\ \sum_{a\in\{0,1\}^m,b\in\{0,1\}^l}
    \qquad \left(\sum_{x\in\{0,1\}^m,y\in\{0,1\}^l}\alpha_{ix}\beta_{(i-1)y}\langle ab|U|xy\rangle\right)|ab\rangle.
\end{align*}
Then we measure $q_1,...,q_m$ in the computational basis, and the outcome $a_i=a_{i1},...,a_{im}\in\{0,1\}^m$ is output through $x_1,...,x_m$, respectively, with probability
\begin{equation}\label{c-prob2}
\begin{aligned}
p(a_i)=\|\langle a_i|U|\varphi_i\rangle|\psi_{i-1}\rangle\|^2
=\sum_{b\in\{0,1\}^l}\left| \sum_{x\in\{0,1\}^m,y\in\{0,1\}^l}\alpha_{ix}\beta_{(i-1)y}\langle a_ib|U|xy\rangle\right|^2.
\end{aligned}
\end{equation}
And variable $p_1,...,p_l$ is in state
\begin{equation}\label{c-state2}
\begin{aligned}
|\psi_i\rangle =\frac{\langle a_i|U|\varphi_i\rangle|\psi_{i-1}\rangle}{\sqrt{p(a_i)}}=\frac{1}{\sqrt{p(a_i)}}\sum_{b\in\{0,1\}^l}
\left(\sum_{x\in\{0,1\}^m,y\in\{0,1\}^l}\alpha_{ix}\beta_{(i-1)y}\langle a_ib|U|xy\rangle\right)|b\rangle,
\end{aligned}
\end{equation}
which is input to the memory elements and stored there.

The relationship between input $|\varphi_i\rangle$, output $a_i$ and state variables $|\psi_i\rangle$ at each step is visualised in Fig. \ref{fig-intuitive}.

\iffalse
\begin{figure}
\centering
\begin{tikzpicture}[
textnode/.style={rectangle, draw=white, fill=white, minimum size=5mm},
squarednode/.style={rectangle, draw=black, fill=white, minimum size=5mm},
]
%Nodes
\node[squarednode] (step1) {Step 1};
\node[textnode] (psi0) [left=0.3cm of step1] {$\ket{\psi_0}$};
\node[textnode] (varphi1) [above=0.3cm of step1] {$\ket{\varphi_1}$};
\node[squarednode] (step2) [right=1cm of step1] {Step 2};
\node[textnode] (varphi2) [above=0.3cm of step2] {$\ket{\varphi_2}$};
\node[squarednode] (step3) [right=1cm of step2] {Step 3};
\node[textnode] (varphi3) [above=0.3cm of step3] {$\ket{\varphi_3}$};
\node[textnode] (psi3) [right=0.3cm of step3] {$\begin{matrix}
a_3 \\ \ket{\psi_3}
\end{matrix}$};

%Lines
\draw[->] (psi0.east) -- (step1.west);
\draw[->] (varphi1.south) -- (step1.north);
\draw[->] (step1.east) -- node[below] {$\ket{\psi_1}$} node[above] {$a_1$} (step2.west);
\draw[->] (varphi2.south) -- (step2.north);
\draw[->] (step2.east) -- node[below] {$\ket{\psi_2}$} node[above] {$a_2$} (step3.west);
\draw[->] (varphi3.south) -- (step3.north);
\draw[->] (step3.east) --  (psi3.west);
\end{tikzpicture}
\caption{An illustration of the execution of a sequential quantum circuit.} \label{fig-intuitive}
\hrulefill \vspace*{4pt}
\end{figure}
\fi

\begin{figure}[!htp]
\centering
\begin{tikzpicture}[
textnode/.style={rectangle, draw=white, fill=white, minimum size=5mm},
squarednode/.style={rectangle, draw=black, fill=white, minimum size=5mm},
]
%Nodes
\node[squarednode] (step1) {Step 1};
\node[textnode] (psi0) [left=0.3cm of step1] {$\ket{\psi_0}$};
\node[textnode] (varphi1) [above=0.3cm of step1] {$\ket{\varphi_1}$};
\node[textnode] (step2) [right=1cm of step1] {\dots};
\node[squarednode] (step3) [right=1cm of step2] {Step $i$};
\node[textnode] (step4) [right=1cm of step3] {\dots};
\node[textnode] (varphi3) [above=0.3cm of step3] {$\ket{\varphi_i}$};
\node[textnode] (a1) [below=0.3cm of step1] {$a_1$};
\node[textnode] (a3) [below=0.3cm of step3] {$a_i$};

%Lines
\draw[->] (psi0.east) -- (step1.west);
\draw[->] (varphi1.south) -- (step1.north);
\draw[->] (step1.east) -- node[below] {$\ket{\psi_1}$} (step2.west);
\draw[->] (step2.east) -- node[below] {$\ket{\psi_{i-1}}$} (step3.west);
\draw[->] (varphi3.south) -- (step3.north);
\draw[->] (step3.east) -- node[below]{$\ket{\psi_i}$} (step4.west);
\draw[->] (step1.south) -- (a1.north);
\draw[->] (step3.south) -- (a3.north);
\end{tikzpicture}
\caption{An illustration of the execution of a sequential quantum circuit.} \label{fig-intuitive}
%\hrulefill \vspace*{4pt}
\end{figure}

\subsection{Equivalence of Sequential Quantum Circuits}

For an unknown sequential quantum circuit $S$ with $m$ input variables and $l$ state variables, the only information that can be obtained is the measurement outcome after each step. Here, the measurement outcome is an $l$-bit binary string.
Regarding the sequential quantum circuits as black-boxes, the only way to distinguish them is to compare their measurement outcomes through experimental statistics. We note that the input states can be selected arbitrarily, but the measurement outcomes are probabilistic.
Two sequential quantum circuits are said to be equivalent if they cannot be distinguished (with a high probability) through these experiments.

Formally, suppose a sequence of input states in $\mathcal{H}_2^{\otimes m}$ is selected to be $\varphi = \ket{\varphi_1}, \dots, \ket{\varphi_K}$. Then the sequential quantum circuit $S$ with initial state $\ket{\psi_0} \in \mathcal{H}^{\otimes l}$ on this input sequence will produce a sequence of measurement outcomes $a = a_1, \dots, a_K$ with probability $p_S \left(a \middle| \varphi, \ket{\psi_0} \right)$ (according to equations (\ref{c-prob2}) and (\ref{c-state2})). A natural definition of equivalence of sequential quantum circuits is given as follows.

\begin{definition} [Equivalence of Sequential Quantum Circuits] \label{def-sqc-equiv}
    Suppose $S_1$ and $S_2$ are two sequential quantum circuits with $m$ input variables and $l_1$ and $l_2$ state variables initialized to $|\psi_0^{(1)}\rangle$ and $|\psi_0^{(2)}\rangle$, respectively. Then $S_1$ and $S_2$ is said to be equivalent, if for every input sequence $\varphi = \ket{\varphi_1}, \dots, \ket{\varphi_K}$ and output sequence $a = a_1, \dots, a_K$, $S_1$ and $S_2$ on input sequence $\varphi$ produce the output sequence $a$ with the same probability, i.e.
    $
        p_{S_1}(a|\varphi, |\psi_0^{(1)}\rangle) = p_{S_2}(a|\varphi, |\psi_0^{(2)}\rangle).
    $
\end{definition}

\begin{remark} Several different notions of equivalence for (combinational) quantum circuits have been introduced in the previous literature (see for example \cite{Via07}). The strongest one requires that the output quantum states must be the same (see Definition \ref{circuit-equivalence} in this paper). The notion of equivalence defined above is weaker than it and only requires that the difference between the output quantum states cannot be detected by the external observation through certain quantum measurements. The applicability of this kind of equivalence is shown by the examples given in Section \ref{sec-case}.

\end{remark}

\section{Quantum Mealy Machines}\label{sec-Mealy}

The Mealy machine \cite{Mealy} is a standard model of classical sequential circuits as a kind of finite-state machine. In this section, we introduce a quantum generalisation of Mealy machines, and show how sequential quantum circuits can be modelled as quantum Mealy machines.

\subsection{Preliminaries and Notations}

A formal definition of a quantum Mealy machine requires more preliminaries from the quantum theory, which we briefly review here (More details can be found in the standard textbook \cite{NC00}). A mixed quantum state is modelled as an ensemble $\{(p_i, \ket{\psi_i})\}$ of pure states, meaning that the system is in state $\ket{\psi_i}$ with probability $p_i$, where $\sum_i p_i = 1$. It can also be described as a density matrix: $
    \rho = \sum_i p_i \ket{\psi_i} \bra{\psi_i}.$
    Let $\mathcal{D}(\mathcal{H})$ denote the set of all density matrices in Hilbert space $\mathcal{H}$.
After a unitary transform $U$ is performed on a quantum system in mixed state $\rho$, the system will be in state $U\rho U^\dag$. A quantum measurement is modelled by a family $M = \{ M_m \}$ of matrices satisfying a normalisation condition $\sum_m M^\dag M_m=I$ (the identity matrix in $\mathcal{H}$), where index $m$ is used to denote different possible outcomes. When we perform $M$ on a quantum system in state $\rho$, outcome $m$ is obtained with probability
$p(m) = \operatorname{tr} (M_m \rho M_m^\dag).$
Right after the measurement,  if the outcome is $m$, the system will be in state $
    \rho' = \frac 1 {p(m)} M_m \rho M_m^\dag.$

Let $A$ and $B$ be two quantum systems. Then the state space of the composite system $AB$ is the tensor product $\h_A\otimes\h_B$ of $A$'s state space $\h_A$ and $B$'s state space $\h_B$. Formally, suppose $\{ \ket{i}_A \}$ and $\{ \ket{j}_B \}$ are orthonormal bases of $\h_A$ and $\h_B$, respectively. Then $\h_A \otimes \h_B = \operatorname{span} \{ \ket{i}_A \otimes \ket{j}_B \}$, the complex vector space with $\{ \ket{i}_A \otimes \ket{j}_B \}$ as an orthonormal basis.
We often use $\ket{i}_A\ket{j}_B$ or simply $\ket{ij}$ to denote $\ket{i}_A \otimes \ket{j}_B$ for short. The inner product of $\ket{ij}$ and $\ket{vw}$ is defined by
    $\braket{ij}{vw} = \braket{i}{v} \braket{j}{w}.$
The state of a subsystem can be described using a reduced density operator. Let $\rho$ be a density operator in $\h_A \otimes \h_B$. Then its reduced density operator for $A$ is given as $
    \rho^A = \operatorname{tr}_B (\rho),$
where $\operatorname{tr}_B$ is a linear map, known as partial trace over $B$, defined by
$\operatorname{tr}_B ( \ket{i}_A\bra{v} \otimes \ket{j}_B\bra{w} ) = \braket{j}{w} \ket{i}_A\bra{v}.$
Suppose $U_A$ and $U_B$ are two operators on $\h_A$ and $\h_B$, respectively. Then the tensor product $U_A \otimes U_B$ is an operator in $\h_a\otimes\h_B$, defined by
$(U_A \otimes U_B) (\ket{i}_A \otimes \ket{j}_B) = U_A\ket{i}_A \otimes U_B\ket{j}_B.$
Moreover, the direct sum of spaces $\h_A$ and $\h_B$ is defined as
    $\h_A \oplus \h_B = \operatorname{span} (\{ \ket{i}_A\}\cup \{\ket{j}_B \})$ provided the two bases are disjoint.
Then the direct sum of operators $U_A$ in $\h_A$ and $U_B$ in $\h_B$ is defined by $(U_A \oplus U_B) \ket{i}_A  = U_A\ket{i}_A$ and  $(U_A \oplus U_B) \ket{j}_B  = U_B\ket{j}_B.$
%Obviously, we have the dimensions: $\dim (\h_A \otimes \h_B)  = \dim \h_A \times \dim \h_B$ and $\dim (\h_A \oplus \h_B)  = \dim \h_A + \dim \h_B.$

\subsection{The Model of Quantum Mealy Machines} \label{sec:def-qmm}

With the preparations in the last subsection, a quantum Mealy machine can be defined as a straightforward abstraction of the model of sequential quantum circuits described in the last section.

\begin{definition}[Quantum Mealy machines] A quantum Mealy machine is a quintuple $\mathcal{M}=\left(\mathcal{H}_\mathit{in}, \mathcal{H}_s, U, M\right),$ where:
$\mathcal{H}_\mathit{in}$ and $\mathcal{H}_s$ are both finite-dimensional Hilbert spaces, called the input space and the state space, respectively; $U$ is a unitary operator on $\mathcal{H}_\mathit{in}\otimes\mathcal{H}_s$; and $M=\{M_a:a\in O\}$ is a quantum measurement in $\mathcal{H}_\mathit{in}\otimes\mathcal{H}_s$.
\iffalse
\begin{itemize}
\item $\mathcal{H}_\mathit{in}$ is a finite-dimensional Hilbert space, called the input space;\item $\mathcal{H}_s$ is also a finite-dimensional Hilbert space, called the state space;
\item $U$ is a unitary operator on $\mathcal{H}_\mathit{in}\otimes\mathcal{H}_s$;\item $M=\{M_a:a\in O\}$ is a quantum measurement in $\mathcal{H}_\mathit{in}\otimes\mathcal{H}_s$.
\end{itemize}
\fi
\end{definition}

The examples of sequential quantum circuits presented in the previous section can be properly specified using the notion of quantum Mealy machine. In our examples, we only use measurements in $H_\mathit{in}$, but for generality, $M$ in the above definition is  allowed to be a measurement in $\mathcal{H}_\mathit{in}\otimes\mathcal{H}_s$. As will be seen in Section \ref{sec-checking}, this generalisation is particularly convenient in the description of equivalence checking of two quantum Mealy machines.

\begin{example} \label{exam-rus-machine}
The RUS implementation circuit of quantum gate $V_3$ in Examples \ref{exam-rus} and \ref{exam-rus+1} can be described as a quantum Mealy machine $\mathcal{M}_\mathit{rus} = (\mathcal{H}_q, \mathcal{H}_p, U, M),$
where: input space $\mathcal{H}_q = \operatorname{span} \{ \ket{0}_q, \ket{1}_q \}$, state space $\mathcal{H}_p = \operatorname{span} \{ \ket{0}_p, \ket{1}_p \}$,
unitary operator $U$ describes the combinational quantum circuit $\mathcal{C}_\mathit{rus}$ defined by equation (\ref{eq-rus}), and
$M = \{ M_0 = \ket {0}_q \bra {0}, M_1 = \ket {1}_q \bra {1} \}$ is the measurement in the computational basis of $\h_q$.
\iffalse
\begin{enumerate}
  \item input space $\mathcal{H}_q = \operatorname{span} \{ \ket{0}_q, \ket{1}_q \}$,
  \item state space $\mathcal{H}_p = \operatorname{span} \{ \ket{0}_p, \ket{1}_p \}$,
  \item unitary operator $U$ describes the combinational quantum circuit $\mathcal{C}_\mathit{rus}$ defined by equation (\ref{eq-rus}),
  \item $M = \{ M_0 = \ket {0}_q \bra {0}, M_1 = \ket {1}_q \bra {1} \}$ is the measurement in the computational basis of $\h_q$.
\end{enumerate}
\fi
\end{example}

\begin{example} \label{exam-walk-machine} The circuit implementing the quantum walk on a $4$-circle in Example \ref{exam-walk} can be described as a quantum Mealy machine $\mathcal{M}_C = (\mathcal{H}_d, \mathcal{H}_c \otimes \mathcal{H}_p, U, M),$
where:
input space $\mathcal{H}_d = \operatorname{span} \{ \ket{0}_d, \ket{1}_d \}$,
state space $\mathcal{H}_c \otimes \mathcal{H}_p$ with $\mathcal{H}_c = \operatorname{span} \{ \ket{0}_c, \ket{1}_c \}$ and $\mathcal{H}_p = \mathcal{H}_{p_1} \otimes \mathcal{H}_{p_2}$, where $\mathcal{H}_{p_i} = \operatorname{span} \{ \ket{0}_{p_i}, \ket{1}_{p_i} \}$ for $i = 1, 2$,
unitary operator $U$ describes the combinational quantum circuit $\mathcal{C}_C$ defined by equation (\ref{eq-walk}), and
$M = \{ M_0 = \ket {0}_d \bra {0}, M_1 = \ket {1}_d \bra {1} \}$ is the measurement in the computational basis of $\h_d$.
\iffalse
\begin{enumerate}
  \item input space $\mathcal{H}_d = \operatorname{span} \{ \ket{0}_d, \ket{1}_d \}$,
  \item state space $\mathcal{H}_c \otimes \mathcal{H}_p$ with $\mathcal{H}_c = \operatorname{span} \{ \ket{0}_c, \ket{1}_c \}$ and $\mathcal{H}_p = \mathcal{H}_{p_1} \otimes \mathcal{H}_{p_2}$, where $\mathcal{H}_{p_i} = \operatorname{span} \{ \ket{0}_{p_i}, \ket{1}_{p_i} \}$ for $i = 1, 2$,
  \item unitary operator $U$ describes the combinational quantum circuit $\mathcal{C}_C$ defined by equation (\ref{eq-walk}),
  \item $M = \{ M_0 = \ket {0}_d \bra {0}, M_1 = \ket {1}_d \bra {1} \}$ is the measurement in the computational basis of $\h_d$.
\end{enumerate}
\fi
\end{example}

Now let us formally describe how a quantum Mealy machine runs. To simplify the description, for each measurement outcome $a\in O$ and input state $\sigma$, we introduce a (super-)operator: for every $\rho$ in $\mathcal{H}_s$,
\begin{equation} \label{eq-superoperator}
\mathcal{E}_{a|\sigma}(\rho)=\operatorname{tr}_{\mathcal{H}_\mathit{in}} \left[M_a U\left(\sigma \otimes\rho\right)U^\dag M_a^\dag\right],
\end{equation}
where $\operatorname{tr}_{\mathcal{H}_\mathit{in}}$ is used to trace out the input subsystem $\mathcal{H}_\mathit{in}$. Then for any input sequence $\sigma_1, \dots , \sigma_i, \dots$ in $\mathcal{H}_\mathit{in}$ and initial (mixed) state $\rho_0$ in $\mathcal{H}_s$, the behaviour of machine $\mathcal{M}$ can be described as follows:
\begin{itemize}
\iffalse
\item \textbf{Step} 1: After receiving the first input, $\mathcal{M}$ is in state $\sigma_1\otimes\rho_0$. Applying unitary $U$ yields $U\left(\sigma_1\otimes\rho_0\right)U^\dag.$ Then we perform measurement $M$ on subsystem $\mathcal{H}_\mathit{in}$, and an outcome $a_1\in O$ is output with probability
\begin{equation}\label{prob1}p(a_1|\sigma_1, \rho_0)=\operatorname{tr}\left[M_{a_1} U\left(\sigma_1\otimes\rho_0\right)U^\dag M_{a_1}^\dag\right].\end{equation}
After that, subsystem $\mathcal{H}_s$ is in state
\begin{equation}\label{state1}\rho_1=\frac{1}{p(a_1|\sigma_1, \rho_0)}
\mathcal{E}_{a_1|\sigma_1}(\rho_0).\end{equation}
\fi
\item \textbf{Step} $i$ $(i\geq 1)$: After receiving the $i$th input, $\mathcal{M}$ is in state $\sigma_i\otimes\rho_{i-1}$. Unitary $U$ transforms it to $U\left(\sigma_i\otimes\rho_{i-1}\right)U^\dag.$ Then we measure subsystem $\mathcal{H}_\mathit{in}$, outcome $a_i\in O$ is output with probability
\begin{equation}\label{prob2}p(a_i|\sigma_i, \rho_{i-1})=\operatorname{tr}\left[M_{a_i}U\left(\sigma_i\otimes\rho_{i-1}\right)U^\dag M_{a_i}^\dag\right],
\end{equation}
and  subsystem $\mathcal{H}_s$ is in state
\begin{equation}\label{state2}\rho_i=\frac{1}{p(a_i|\sigma_i, \rho_{i-1})}
\mathcal{E}_{a_i|\sigma_i}(\rho_{i-1}).\end{equation}
\end{itemize}
It is easy to see that equations  (\ref{c-prob2}) and (\ref{c-state2}) are special cases of (\ref{prob2}) and (\ref{state2}), respectively.

The application of an input sequence is called an \textit{experiment} on the machine. More explicitly, an input sequence $\pi=\sigma_1, \dots, \sigma_k$
%of length $k$
and an initial state $\rho_0$ induce a probability distribution $p(\cdot |\pi,\rho_0)$ over output sequences $O^k$ of length $k$: for every $a=a_1,...,a_k\in O^k$,
$p(a|\pi,\rho_0)=\prod_{i=1}^k p(a_i|\sigma_i, \rho_{i-1})$,
where $p(a_i|\sigma_i, \rho_{i-1})$ is defined by equation (\ref{prob2}).

\begin{remark} The reader should have noticed that in the above description of running a quantum Mealy machine, the initial state in $\mathcal{H}_s$ is taken as a mixed state $\rho_0$ rather than a pure state $|\psi_0\rangle$. The reason behind this design decision is that even if the machine starts in a pure initial state $|\psi_0\rangle$, the subsequent states $\rho_1,...,\rho_i,...$ can still be mixed states because they are obtained by tracing out the input subsystem $\mathcal{H}_\mathit{in}$ (see equation~(\ref{state2})).
\end{remark}

%(2) As is well known, the output of a Mealy machine depends on both the input and the machine's (internal) state, but the output of a Moore machine depends only on the machine's state. At the first glance, the model defined above is more like a quantum generalisation of Moore machines because the output comes directly from measurement $M$ (see Eqs. (\ref{prob1}) and (\ref{prob2})). But we argue that the model is actually a quantum generalisation of Mealy machines: after unitary operator $U$ acts on $\mathcal{H}_\mathit{in}\otimes\mathcal{H}_s$, the input state in $\mathcal{H}_\mathit{in}$ is entangled with the machine's state in $\mathcal{H}_s$. Then measurement $M$ is performed on $\mathcal{H}_\mathit{in}$ (but not $\mathcal{H}_s$). Consequently, the outcome of $M$ depends directly on the input state in $\mathcal{H}_\mathit{in}$ and (less explicitly) on the machine's state in $\mathcal{H}_s$.

\begin{remark}\label{remark-difference-1} At the first glance, the notion of quantum Mealy machine defined above is similar to the one defined in \cite{Wang19},  but they are actually very different. The one defined above is designed for modelling sequential quantum circuits, whereas the one in \cite{Wang19} was used for modelling combinational quantum circuits. This difference between the design purposes leads to the differences between the components of machines. First, the input to the machine in  \cite{Wang19} is a string over a finite alphabet (that is, a sequence of classical states),  while that of the machine in this paper is a sequence of quantum states in a finite-dimensional Hilbert space. In an execution, the input of the machine in \cite{Wang19}  indicates which unitary operator is performed
%(and the original input remains unchanged),
while  the input to the machine in this paper  coherently gets involved in the computation and possibly becomes entangled with the internal state of the machine.
Thus, to fetch information, we must perform measurements on both input and state Hilbert spaces, rather than the only state space as in \cite{Wang19}. Thus we need to use partial trace (see equation (\ref{eq-superoperator})) to trace out the input space, while \cite{Wang19} does not. Furthermore, the intermediate states during the execution are mixed states in both \cite{Wang19} and this paper, but the reasons are different. In \cite{Wang19}, the initial (internal) state is allowed to be a mixed state and thus the subsequent intermediate states would be mixed states too. On the contrary, the intermediate mixed states are inevitable in the  machine studied in this paper. Even if the initial internal state is a pure state, the intermediate states could still be mixed states due to the possible entanglement between the input and internal states before measurement.
\end{remark}

\subsection{Equivalence of Quantum Mealy Machines}

To conclude this section, we introduce the key notion of equivalence for quantum Mealy machines.

\begin{definition}[Equivalence of quantum Mealy machines]\label{equivalence} Let $\mathcal{M}_i=\left(\mathcal{H}_\mathit{in}, \mathcal{H}^{(i)}_s, U_i, M_i\right)$ $(i=1,2)$ be two quantum Mealy machines with the same input space $\mathcal{H}_\mathit{in}$ and the same outputs $O$ (i.e. measurements $M_i=\left\{M_a^{(i)}:a\in O\right\}$ for $i=1,2$), let $\rho_i\in\mathcal{D}(\mathcal{H}_s^{(i)})$ $(i=1,2)$, $V \subseteq \mathcal{D}(\mathcal{H}_\mathit{in})$, and let $K$ be a positive integer. Then:\begin{enumerate}\item $\mathcal{M}_1$ with initial state $\rho_1$ and $\mathcal{M}_2$ with initial state $\rho_2$ are (functionally) equivalent, written: $(\mathcal{M}_1,\rho_1)\sim (\mathcal{M}_2,\rho_2),$ if for any input sequence $\pi$ and for any $a\in O^{|\pi|}$, \begin{equation}\label{def-se}p\left(a|\pi,\rho_1\right)=p\left(a|\pi,\rho_2\right).
\end{equation} Note that the probability in the left-hand side of the above equation is defined in machine $\mathcal{M}_1$ with $\rho_1$ as its initial state, and the probability in the right-hand side is given in $\mathcal{M}_2$ with initial state $\rho_2$.
\item $\mathcal{M}_1$ with $\rho_1$ and $\mathcal{M}_2$ with $\rho_2$ are $V$-equivalent, written: $(\mathcal{M}_1,\rho_1)\sim_{V} (\mathcal{M}_2,\rho_2),$ if equation (\ref{def-se}) holds for all input sequences $\pi$ in $V$.
\item $\mathcal{M}_1$ with $\rho_1$ and $\mathcal{M}_2$ with $\rho_2$ are $K$-equivalent, written: $(\mathcal{M}_1,\rho_1)\sim_K (\mathcal{M}_2,\rho_2),$ if equation (\ref{def-se}) holds for all input sequences $\pi$ with $|\pi|\leq K$.
\item $\mathcal{M}_1$ with $\rho_1$ and $\mathcal{M}_2$ with $\rho_2$ are $(V, K)$-equivalent, written: $(\mathcal{M}_1,\rho_1)\sim_{V,K} (\mathcal{M}_2,\rho_2),$ if equation (\ref{def-se}) holds for all input sequences $\pi$ in $V$ with $|\pi|\leq K$.
\end{enumerate}
\end{definition}

Intuitively, $(\mathcal{M}_1,\rho_1)\sim (\mathcal{M}_2,\rho_2)$ means the probability distributions of the outputs of the two machines $\mathcal{M}_1$ and $\mathcal{M}_2$ are the same for any sequence of inputs, but $V$-equivalence only requires the distributions of outputs are the same for all inputs from a given range $V$, and $K$-equivalence only requires the distributions are the same for all input and output sequences of length not greater than $K$.
Whenever $\mathcal{M}_1$ and $\mathcal{M}_2$ are the same and $(\mathcal{M}_1,\rho_1)\sim (\mathcal{M}_2,\rho_2)$ (or $(\mathcal{M}_1,\rho_1)\sim_V (\mathcal{M}_2,\rho_2)$, $(\mathcal{M}_1,\rho_1)\sim_K (\mathcal{M}_2,\rho_2)$, $(\mathcal{M}_1,\rho_1)\sim_{V,K} (\mathcal{M}_2,\rho_2)$), we simply say that $\rho_1$ and $\rho_2$ are equivalent (resp. $V$-equivalent, $K$-equivalent, $(V,K)$-equivalent) and write $\rho_1\sim \rho_2$ (resp. $\rho_1\sim_V \rho_2$, $\rho_1\sim_K \rho_2$, $\rho_1\sim_{V,K} \rho_2$). When two pure states $\ket{\psi_1}$ and $\ket{\psi_2}$ are equivalent, i.e. $\ket{\psi_1}\bra{\psi_1} \sim \ket{\psi_2}\bra{\psi_2}$, we simply write $\ket{\psi_1} \sim \ket{\psi_2}$.

%To conclude this section, let us see how the equivalence of quantum Mealy machines is linked to the equivalence of the sequential quantum circuits modeled by these machines.
To conclude this section, let us see how the equivalence of quantum Mealy machines is linked to that of sequential quantum circuits.
Suppose $S$ is a sequential quantum circuit with $m$ input variables and $l$ state variables. Then $S$ can be modelled by a quantum Mealy machine $\mathcal{M}(S) = (\mathcal{H}_\mathit{in}, \mathcal{H}_s, U, M)$, where $\mathcal{H}_\mathit{in} = \mathcal{H}_2^{\otimes m}$, $\mathcal{H}_\mathit{s} = \mathcal{H}_2^{\otimes l}$, $U$ is the combinational quantum circuit of $S$, and $M = \{M_a = \ket{a}_\mathit{in}\bra{a}: a \in \{0, 1\}^m\}$ is the measurement. %in the computational basis of $\mathcal{H}_\mathit{in}$.
Then we have:

\begin{theorem}
    Suppose $S_1$ and $S_2$ are two sequential quantum circuits as Definition \ref{def-sqc-equiv}. Then $S_1$ and $S_2$ are equivalent if and only if $(\mathcal{M}(S_1), |\psi_0^{(1)}\rangle) \sim (\mathcal{M}(S_2), |\psi_0^{(2)}\rangle)$.
\end{theorem}
\textit{Proof}: It is straightforward from the definitions.
\hfill $\blacksquare$

The above theorem provides a basis for using quantum Mealy machines in equivalence checking of sequential quantum circuits. To see this more clearly, some examples are provided in Section \ref{sec:def-qmm}.

\section{Equivalence Checking}\label{sec-checking}

In this section, we present the main theoretical result of this paper. In particular, we develop an algorithm for checking equivalence of two quantum Mealy machines with initial states.

\subsection{Direct Sum of Quantum Mealy Machines}

The basic idea of our algorithm is a reduction from equivalence checking of two quantum Mealy machines with initial states to equivalence checking of two states in their direct sum, which is defined in the following:

\begin{definition}[Direct sum of quantum Mealy machines]\label{def:direct-sum}
    Let $\mathcal{M}_i$ $(i=1,2)$ be the same as in Definition \ref{equivalence}. Then the direct sum of $\mathcal{M}_1$ and $\mathcal{M}_2$ is quantum Mealy machine
        $$\mathcal{M}_1 \oplus \mathcal{M}_2 = \left(\mathcal{H}_\mathit{in}, \mathcal{H}^{(1)}_s \oplus \mathcal{H}^{(2)}_s, U_1 \oplus U_2, M\right),$$
    where $\oplus$ in the right-hand side of the above equation stands for direct sum of two Hilbert spaces or two operators, and measurement $M=\{M_a: a\in O\}$ with $M_a = M_a^{(1)} \oplus M_a^{(2)}$ for every $a \in O$.\end{definition}

    To see that $\mathcal{M}_1 \oplus \mathcal{M}_2$ is well-defined as a quantum Mealy machine, it suffices to note that for each $a\in O$, $M_a = M_a^{(1)} \oplus M_a^{(2)}$ is a matrix in $(\mathcal{H}_\mathit{in}\otimes\mathcal{H}^{(1)}_s) \oplus (\mathcal{H}_\mathit{in}\otimes\mathcal{H}^{(2)})\cong
\mathcal{H}_\mathit{in}\otimes (\mathcal{H}^{(1)}_s \oplus \mathcal{H}^{(2)})$, and normalisation condition $\sum_{a\in O}M_a^\dag M_a=I$ (the identity matrix in $\mathcal{H}_\mathit{in}\otimes(\mathcal{H}^{(1)}_s \oplus \mathcal{H}^{(2)})$) is satisfied.

\begin{remark}\label{remark-difference-2}
In \cite{Wang19}, since the input is a sequence of classical states (that is, a string over a finite alphabet), the direct sum of two machines can be defined straightforwardly. In contrast, as discussed in Remark \ref{remark-difference-1}, the input to the machine considered in this paper is a sequence of quantum states that can be entangled with the internal states. Then a major difficulty in defining the sum of two machines is to make the input entangled with the internal states of the two machines ``simultanueously''. To resolve this issue, we define the measurement operators of the sum to be on Hilbert space $\mathcal{H}_\mathit{in} \otimes (\mathcal{H}_s^{(1)} \oplus \mathcal{H}_s^{(2)})$, which keep the behaviour of each machine as if they did not interfere with each other. In comparison, the measurement operators of the sum machine defined in  \cite{Wang19} are on $\mathcal{H}_s^{(1)} \oplus \mathcal{H}_s^{(2)}$.
%Though the measurements are only performed on input variables in sequential quantum circuits, it would be helpful to make some measurements on internal states or ignore some measurements on input states in some general cases. Moreover, technically, the general definition of measurements will help to define the direct sum of two quantum Mealy machines conveniently (see Definition \ref{def:direct-sum}).
\end{remark}

\subsection{Main Theorem}

There are two major difficulties in designing an algorithm for equivalence checking of quantum circuits. As in checking classical circuits, it is desirable to find an upper bound of the length of the needed input sequences. On the other hand, the input state space of a classical circuit is finite; in contrast, the input state of a quantum circuit is a continuum. These difficulties are resolved by the following theorem showing that equivalence checking of quantum Mealy machines can be done using only input sequences taken from some given basis and of length quadratic in the dimensions of the state Hilbert spaces of the machines.
Surprisingly, the length of checking input sequences is independent of the dimension of the input space.

\begin{theorem} \label{thm-eq}
    Let $\mathcal{M}_i=\left(\mathcal{H}_\mathit{in}, \mathcal{H}^{(i)}_s, U_i, M_i\right)$ $(i=1,2)$ be two quantum Mealy machines with the same input space $\mathcal{H}_\mathit{in}$ and the same outputs $O$ (i.e. measurements $M_i=\left\{M_a^{(i)}:a\in O\right\}$ for $i=1,2$), let $\rho_i\in\mathcal{D}(\mathcal{H}_s^{(i)})$ $(i=1,2)$, and let $V$ be a subspace of $\mathcal{D}(\mathcal{H}_\mathit{in})$. Then the following two statements are equivalent:
    \begin{enumerate}
      \item $(\mathcal{M}_1,\rho_1) \sim_V (\mathcal{M}_2,\rho_2)$.
      \item $\rho_1\sim_{B, d_1^2+d_2^2-1}\rho_2$ (in $\mathcal{M}_1 \oplus \mathcal{M}_2$), where $B$ is a basis of $V$, and
      $d_i = \dim \mathcal{H}_s^{(i)}$ for $i=1,2$.
       \end{enumerate}
\end{theorem}

\textit{Proof}: Let us first outline the basic idea of the proof. We are going to prove the theorem in three steps:\begin{enumerate}\item[(i)] $(\mathcal{M}_1,\rho_1)\sim_V (\mathcal{M}_2,\rho_2)$ iff $\rho_1\sim_V \rho_2$ in $\mathcal{M}_1\oplus \mathcal{M}_2$; \item[(ii)] $\rho_1\sim_B \rho_2$ in $\mathcal{M}$ iff $\rho_1\sim_{\operatorname{span} B} \rho_2$, where $\operatorname{span} B$ is the subspace spanned by $B$; \item[(iii)] $\rho_1\sim_B \rho_2$ iff $\rho_1\sim_{B,d^2-1} \rho_2$, where $d$ is the dimension of the state Hilbert space of $\mathcal{M}$.\end{enumerate}

The first two steps are intuitive and we put their tedious details in Appendix \ref{app:thm-eq}. %in the full version \cite{WLY18} of this paper. %into Appendix \ref{appendix1}.
To prove step (iii),
let $\rho$ be a density operator and $B \subseteq \mathcal{D}(\mathcal{H}_\mathit{in})$ be a finite set with $\operatorname{span} B = \operatorname{span} V$. By step (ii), we have $\rho_1 \sim_V \rho_2$ iff $\rho_1 \sim_B \rho_2$. Let
$
    D(\rho, m) = \{ \mathcal{E}_{a|\pi}(\rho): |\pi| \leq m, a \in O^{|\pi|}, \pi_i \in B \}
$.
We observe that $D(\rho, m) \subseteq D(\rho, m+1)$, and thus $\operatorname{span} D(\rho, m) \subseteq \operatorname{span} D(\rho, m+1)$ for every $m \in \mathbb{N}$.

{\vskip 4pt}

\textbf{Claim}. If for some $m \in \mathbb{N}$, $\operatorname{span} D(\rho, m) = \operatorname{span} D(\rho, m+1)$, then $\operatorname{span} D(\rho, m) \subseteq \operatorname{span} D(\rho, m+\delta)$ for all $\delta \in \mathbb{N}$.

We prove this claim by induction. It is trivial for the case $\delta = 1$. Suppose it holds for $\delta > 0$. Then for every $t \leq m+\delta+1$ and $\mathcal{E}_{a_1\dots a_t|\sigma_1 \dots \sigma_t}(\rho) \in D(\rho, m+\delta+1)$, we have: $\mathcal{E}_{a_1\dots a_{t-1}|\sigma_1 \dots \sigma_{t-1}}(\rho) \in \operatorname{span} D(\rho, m+\delta) = \operatorname{span} D(\rho, m)$ by the induction hypothesis, and
\begin{align*}
    \mathcal{E}_{a_1\dots a_t|\sigma_1 \dots \sigma_t}(\rho)
    & = \mathcal{E}_{a_t|\sigma_t}(\mathcal{E}_{a_1\dots a_{t-1}|\sigma_1 \dots \sigma_{t-1}}(\rho)) \\
    & \in \mathcal{E}_{a_t|\sigma_t}(\operatorname{span} D(\rho, m+\delta)) \\
    & = \mathcal{E}_{a_t|\sigma_t}(\operatorname{span} D(\rho, m)) \\
    & \subseteq \operatorname{span} D(\rho, m+1)
     = \operatorname{span} D(\rho, m),
\end{align*}
which implies that $\operatorname{span} D(\rho, m+\delta+1) = \operatorname{span} D(\rho, m)$, that is, the claim is true for $\delta+1$.

Now we notice that the claim implies: $\operatorname{span} D(\rho, d^2-1) \supseteq \operatorname{span} D(\rho, m)$ for every $m \in \mathbb{N}$ because $\dim \operatorname{span} D(\rho, m) \leq d^2$ and $\dim \operatorname{span} D(\rho, 0) = 1$.
Suppose $\rho_1$ and $\rho_2$ are $(B, d^2-1)$-equivalent. Then $\operatorname{tr}(\varrho) = 0$ for any $\varrho \in D(\rho, d^2-1)$, where $\rho = \rho_1 - \rho_2$. On the other hand, for every input sequence $\pi$ and $a \in O^{|\pi|}$, we have: $\mathcal{E}_{a|\pi}(\rho) \in D(\rho, |\pi|) \subseteq \operatorname{span} D(\rho, d^2-1).$ Thus,
\iffalse
\begin{align*}
& p(a|\pi, \rho_1)-p(a|\pi, \rho_2)
 = \operatorname{tr}\left[\mathcal{E}_{a|\pi}(\rho_1)\right] - \operatorname{tr}\left[\mathcal{E}_{a|\pi}(\rho_2)\right] \\
 & \qquad \qquad = \operatorname{tr}\left[\mathcal{E}_{a|\pi}(\rho_1-\rho_2)\right]
 = \operatorname{tr}\left[\mathcal{E}_{a|\pi}(\rho)\right]
 = 0,
\end{align*}
\fi
$
 p(a|\pi, \rho_1)-p(a|\pi, \rho_2)
 = \operatorname{tr}\left[\mathcal{E}_{a|\pi}(\rho_1)\right] - \operatorname{tr}\left[\mathcal{E}_{a|\pi}(\rho_2)\right]
 = \operatorname{tr}\left[\mathcal{E}_{a|\pi}(\rho_1-\rho_2)\right]
 = \operatorname{tr}\left[\mathcal{E}_{a|\pi}(\rho)\right]
 = 0,
$
and this completes the proof of step (iii).

Finally, we can prove Theorem \ref{thm-eq} by gluing the three steps. Our strategy is to show that clauses 1) and 2) of Theorem \ref{thm-eq} and the following two statements are all equivalent:
    \begin{enumerate}
      %\item $(\mathcal{M}_1, \rho_1)\sim_V (\mathcal{M}_2,\rho_2)$;
      \item[a)] $\rho_1 \sim_B \rho_2$ (in $\mathcal{M}_1 \oplus \mathcal{M}_2$), where $B$ is a basis of $V$;
      %\item $\rho_1 \sim_{B,d_1^2+d_2^2-1} \rho_2$ (in $\mathcal{M}_1 \oplus \mathcal{M}_2$), where $d_i = \dim \mathcal{H}_s^{(i)}$ for $i = 1, 2$;
      \item[b)] $(\mathcal{M}_1, \rho_1)\sim_{d_1^2+d_2^2-1} (\mathcal{M}_2,\rho_2)$.
    \end{enumerate}

a) $\Longrightarrow$ 2) and 2) $\Longleftrightarrow$ b) are trivial. By steps (i) and (ii), it holds that 1) $\Longleftrightarrow$ a). Therefore,
we only need to prove that 2) $\Longrightarrow$ a).
Let $\mathcal{E}_{a|\pi}^{(1)}, \mathcal{E}_{a|\pi}^{(2)}$ and $\mathcal{E}_{a|\pi}$ be the super-operator defined in $\mathcal{M}_1, \mathcal{M}_2$ and $\mathcal{M} = \mathcal{M}_1 \oplus \mathcal{M}_2$ by equation (\ref{eq-superoperator}) and generalised by equation (\ref{eq-composition}), respectively, and let $\rho = \rho_1 \oplus (-\rho_2)$. Similar to the proof of step (iii), we define
$
D(\rho, m) = \{ \mathcal{E}_{a|\pi}(\rho): |\pi| \leq m, a \in O^{|\pi|}, \pi_i \in B \}.
$
From the definition of $\mathcal{M}_1 \oplus \mathcal{M}_2$, it holds that
$\mathcal{E}_{a|\pi}(\rho) = \mathcal{E}_{a|\pi}^{(1)}(\rho_1) \oplus \left(-\mathcal{E}_{a|\pi}^{(2)}(\rho_2)\right).$
Thus, we conclude that
$
D(\rho, m) \subseteq \mathcal{D}(\mathcal{H}_s^{(1)}) \oplus \mathcal{D}(\mathcal{H}_s^{(2)}),
$
which implies:
\[
\dim \operatorname{span} D(\rho, m) \leq \dim \mathcal{D}(\mathcal{H}_s^{(1)}) + \dim \mathcal{D}(\mathcal{H}_s^{(2)}) \leq d_1^2+d_2^2.
\]
With the above inequality, we obtain that $\operatorname{span} D(\rho, d_1^2+d_2^2-1) \supseteq \operatorname{span} D(\rho, m)$ for every $m \in \mathbb{N}$. Then the remaining part of the proof can be carried out in the same way as the proof of step (iii).
\hfill $\blacksquare$

\begin{remark} The reader might noticed that high-level structure of the above proof is similar to the proof of the main result in \cite{Wang19}.
This reason is that both of the proofs are quantum generalization of the  technique of \cite{Tze92} for probabilistic automata, which is now a standard and basic technique in automata theory. However, the difference between the machines studied in \cite{Wang19} and this paper as well as the difference between their direct sums (see Remarks \ref{remark-difference-1} and \ref{remark-difference-2}) leads to some differences in the proof details. In particular, the input to the  quantum automata considered in the previous literature including \cite{Wang19} is taken from a finite alphabet and thus can be easily enumerated. To reduce the equivalence of two machines to a property that is easy to check, an observation in the above proof is that only a basis of the input Hilbert space matters, thanks to the linearity of quantum super-operators.
\end{remark}

Clause 2) of Theorem \ref{thm-eq} indicates that the length of input sequences for checking equivalence of two quantum Mealy machines is required to be quadratic in the dimensions of their state spaces. Essentially, the quadratic length comes from the dimensions of the spaces of mixed states, which occur inevitably even if the input is a pure state, because measurements can be performed at the middle of a computation. Intuitively, the space spanned by the mixed states that are generated by the input and output sequences of length not greater than $l$ is of dimension not less than $\min\{l+1, d\}$, where $d = d_1^2+d_2^2$ is the dimension of the space of all possible mixed states in the state Hilbert space
(This choice of $d$ was also used in \cite{LQ09}. Here we derive it naturally from the dimension of density operators).
Therefore, $l$ is required to set to be at least $d-1$ in order to make that the spanned space coincides with the space of all possible mixed states.
On the other hand, it is not enough to require input sequences to be less than or equal to the dimension of the state Hilbert space, as already pointed out in \cite{Wang19}.

In Theorem \ref{thm-eq}, the choice of the basis $B$ of subspace $V$ can be arbitrary. For the most common case that $V = \mathcal{D}(\mathcal{H}_\mathit{in})$,   there is a convenient basis of the space of all density operators that consists of only stabilizer states \cite{Gay11}. As shown in~\cite{Gay18}, stabilizer formalism can greatly simplify the equivalence checking of some special quantum systems.

\subsection{Algorithm}

An algorithm for checking equivalence of two quantum Mealy machines can be directly derived from Theorem \ref{thm-eq} by enumerating all possible input sequences in $B$ within length $d_1^2+d_2^2-1$, but its complexity is $d^{\mathcal{O}(d_1^2+d_2^2)}$, where $d = \dim V$, exponential in the dimensions $d_1$ and $d_2$ of state Hilbert spaces of the machines. Here, we are able to develop
%a much more
a more efficient algorithm with a time complexity polynomial in $d_1$ and $d_2$.

\subsubsection{Description of the Algorithm}
We assume that $B$ is a basis of $V$, which is a given subspace of $\mathcal{D}(\mathcal{H}_\mathit{in})$, and $O$ is the set of outputs.
Note that the choice of $B$ is arbitrary; that is, $B$ can be any basis of $V$. In particular, when $V = \mathcal{D}(\mathcal{H}_\mathit{in})$, we can choose $B$ to be the computational basis.
In this algorithm, the queue is maintained to be monotonic in an admissible order on the set $(B \times O)^*$ that first compares the length and then the lexicographical order. That is, $u < v$ for two different $u, v \in (B \times O)^*$, if $\abs{u} < \abs{v}$, or $\abs{u} = \abs{v}$ and $u$ has priority over $v$ in lexicographical order. As we will see in the complexity analysis given in Section \ref{sec-complexity}, the admissible order in the queue enables the algorithm to terminate in a time polynomial in $d_1$ and $d_2$. For input sequence $\pi=\sigma_1, \dots, \sigma_k$ and output sequence $a=a_1,\dots,a_k$, we define:
\begin{equation}\label{eq-composition}
\mathcal{E}_{a|\pi}=\mathcal{E}_{a_k|\sigma_k}\circ\cdots\circ\mathcal{E}_{a_1|\sigma_1}
\end{equation}
(composition of (super-)operators), where super-operators $\mathcal{E}_{a_i|\sigma_i}$ are defined in $\mathcal{M}_1 \oplus \mathcal{M}_2$. In case of $k = 0$ (i.e. the input sequence is empty), $\mathcal{E}_{a|\pi}$ is defined to be $\mathcal{I}$, the identity (super-)operator, which transforms any density operator to itself, i.e. $\mathcal{I}(\rho) = \rho$ for every $\rho$.

The algorithm is based on Lex-BFS (lexicographical breadth-first search). A queue $Q$ is used to maintain possible candidates of $(\pi, a)$ in the admissible order discussed above, and a set $\mathfrak{B}$ is used to collect a basis of $\mathcal{E}_{a|\pi}(\rho)$ for all $\pi$ and $a$ chosen from the queue $Q$ in the execution of the algorithm, where $\rho = \rho_1 \oplus (-\rho_2)$.
\begin{enumerate}
  \item Initially, a pair $(\pi = \epsilon, a = \epsilon)$ of empty sequences is pushed into $Q$, and the set $\mathfrak{B}$ is set to be empty.

  \item Whenever the queue is not empty, pop the front element $(\pi, a)$ of the queue, and check whether $\mathcal{E}_{a|\pi}(\rho) \in \operatorname{span}\mathfrak{B}$ or not. If not, add $\mathcal{E}_{a|\pi}(\rho)$ into $\mathfrak{B}$, and push $(\pi\sigma, ax)$ into $Q$ for all $\sigma \in B$ and $x \in O$.

  \item When the queue is empty, check whether $\operatorname{tr}(\varrho) = 0$ for all $\varrho \in \mathfrak{B}$ or not. If yes, then claim that $\rho_1 \sim_V \rho_2$, and $\rho_1 \not\sim_V \rho_2$ otherwise. It should be noted that $\operatorname{tr}\left(\mathcal{E}_{a|\pi}(\rho)\right) = 0$ implies the equivalence of $\mathcal{M}_1$ and $\mathcal{M}_2$ on input sequence $\pi$ and output sequence $a$.
\end{enumerate}

The algorithm described above is formalised in Algorithm~\ref{algo}.

    \begin{algorithm}[!htp]
        \caption{Checking equivalence of quantum Mealy machines.}
        \label{algo}
        \begin{algorithmic}[1]
        \Require $\mathcal{M}_i$, $\rho_i$ $(i=1,2)$ and $B$ the same as in Theorem \ref{thm-eq}.
        \Ensure Whether $(\mathcal{M}_1, \rho_1) \sim_V (\mathcal{M}_2, \rho_2)$ or not.

        \State $\rho \gets \rho_1 \oplus (-\rho_2)$.
        \State $\mathfrak{B} \gets \emptyset$.
        \State Let $Q$ be an empty queue and push $(\epsilon, \epsilon)$ into $Q$.
        \While {$Q$ is not empty}
            \State Pop the front element $(\pi, a)$ of $Q$.
            \If {$\mathcal{E}_{a|\pi}(\rho) \notin \operatorname{span} \mathfrak{B}$}
                \State Add $\mathcal{E}_{a|\pi}(\rho)$ into $\mathfrak{B}$.
                \State Push $(\pi \sigma, a x)$ into $Q$ for $\sigma \in B$ and $x \in O$.
            \EndIf
        \EndWhile
        \State \Return \textbf{true} if $\operatorname{tr}(\varrho) = 0$ for every $\varrho \in \mathfrak{B}$, and \textbf{false} otherwise.

        \end{algorithmic}
    \end{algorithm}

    \subsubsection{Correctness and Complexity} \label{sec-complexity}

    Correctness of Algorithm \ref{algo} can be proved using Theorem \ref{thm-eq}. For readability, the detailed proof is put in Appendix \ref{app:correctness}.
    %the full version \cite{WLY18} of this paper.
    %postponed into Appendix \ref{appendix-correctness-of-algorithm}.

    The complexity of Algorithm \ref{algo} is analysed as follows. All mixed states are supposed to be stored as matrices with appropriate dimensions, and all quantum operations are then seen as the corresponding matrix operations. Since $\mathfrak{B}$ is maintained to be linearly independent, $\abs{\mathfrak{B}} \leq d_1^2+d_2^2$, where $d_i = \dim \mathcal{H}_s^{(i)}$ for $i = 1, 2$.
    Thus, there are at most $\abs{B} \abs{O} \abs{\mathfrak{B}} \leq d_V(d_1^2+d_2^2) \abs{O}$ elements that are pushed into the queue, where $d_V = \dim V = \abs{B}$. For each element $\mathcal{E}_{a|\pi}(\rho)$ in the queue, it requires to check whether $\mathcal{E}_{a|\pi}(\rho) \in \operatorname{span} \mathfrak{B}$, which needs $\mathcal{O}((d_1^2+d_2^2)^3)$ time (for example, using Gaussian elimination).
    Here, we also need $\mathcal{O}(d_\mathit{in}^3 (d_1^3+d_2^3))$ (by matrix multiplication) to compute $\mathcal{E}_{a|\pi}(\rho)$ based on previous calculations, where $d_\mathit{in} = \dim \mathcal{H}_{\mathit{in}}$.
    Therefore, the total time complexity is $\mathcal{O} (\abs{O} d_V (d_1^2+d_2^2)^4) + \mathcal{O}(\abs{O} d_V d_\mathit{in}^3 (d_1^3+d_2^3)(d_1^2+d_2^2)) = \mathcal{O}(\abs{O}d_V d^5 (d^3 + d_\mathit{in}^3) )$, where $d = d_1+d_2$.
    On the other hand, only the mixed states in $\mathfrak{B}$ are required to be stored, while other mixed states are calculated only when needed. For example, $\mathcal{E}_{\pi\sigma|ax}(\rho)$ can be calculated by $\mathcal{E}_{\pi\sigma|ax}(\rho) = \mathcal{E}_{\sigma|x}\left(\mathcal{E}_{\pi|a}(\rho)\right)$. Therefore, the space complexity is $\mathcal{O}(\abs{\mathfrak{B}}^2) = \mathcal{O} ((d_1^2+d_2^2)^2)$.
    Formally, we have:
    \begin{theorem} \label{thm-complexity}
        There is an algorithm with time complexity $\mathcal{O}(\abs{O}d_V d^5 (d^3 + d_\mathit{in}^3))$ and space complexity $\mathcal{O} ((d_1^2+d_2^2)^2)$ that, given $\mathcal{M}_i$, $\rho_i$ $(i = 1, 2)$ and $V$ be the same as in Theorem \ref{thm-eq}, checks whether $\rho_1 \sim_V \rho_2$ or not, where $d = d_1 + d_2$, $d_i = \dim \mathcal{H}_s^{(i)}$ for $i = 1, 2$ and $d_V = \dim V$.
    \end{theorem}
   % \begin{cor}
      %  There is a $\mathcal{O} (\abs{O} d^2 (d_1^2+d_2^2)^4)$ algorithm that, given $\mathcal{M}_i$ and $\rho_i$ $(i = 1, 2)$ be the same as in Theorem \ref{thm-eq}, checks whether $\rho_1 \sim \rho_2$ or not, where $d_i = \dim \mathcal{H}_s^{(i)}$ for $i = 1, 2$ and $d = \dim \mathcal{H}_\mathit{in}$.
   % \end{cor}

    A careful comparison between the complexity of our algorithm for checking equivalence of sequential quantum circuits and the complexity of the known best algorithm for checking classical circuits should be helpful for understanding the difference between the classical and quantum cases.
   To check equivalence of two sequential quantum circuits with $m$ input qubit variables and $l_1$ and $l_2$ state qubit variables, respectively, we need to phrase them into the model of our quantum Mealy machine with $\abs{O} = 2^m$, $d_V = \dim \mathcal{D}(\mathcal{H}_\mathit{in}) = 2^{2m}$, $d_i = 2^{l_i}$ for $i = 1, 2$ and $d = d_1 + d_2 = 2^l$, where $l = \max\{l_1,l_2\}$. The overall complexity is then $\mathcal{O}(2^{3m+5l}(2^{3m}+2^{3l}))$, as given by Theorem \ref{thm-complexity}.
   On the other hand, the state space traversal based algorithms have been extensively adopted for checking equivalence of classical sequential classical circuits; see for example \cite{Jha} and \cite{Mol04}. Although several industrial improvements are proposed, for example \cite{Eij06} and \cite{Sav14}, the worst case complexity remains unimproved.
Table \ref{table2} gives a comparison between the complexities of checking equivalence of sequential quantum circuits and classical ones with the same numbers of input and state (qubit/Boolean) variables.
From there, we see that both algorithms have time complexities exponential in the number of (qu)bits.

    \begin{table}[!htp]
    \centering
    \caption{Complexity comparison between quantum and classical.}
    \begin{tabular} {|c|c|c|}
    \hline
     & Quantum & Classical \\
    \hline
    Complexity & $\mathcal{O}(2^{3m+5l}(2^{3m}+2^{3l}))$ & $\mathcal{O}(2^{m+l})$ \\
    \hline
    \end{tabular}
    \begin{tablenotes}
      \small
      \item This table shows the complexity of checking equivalence of two sequential quantum (resp. classical) circuits with $m$ input qubit (resp. bit) variables and $l$ state qubit (resp. bit) variables, respectively.
    \end{tablenotes}
    \label{table2}
\end{table}

\section{Case Studies}\label{sec-case}

To show the utility of the formal model defined in this paper and further illustrate Algorithm \ref{algo}, we present several case studies in this section.

\subsection{Repeat-until-Success Implementation}

We first consider the repeat-until-success implementation example (Example \ref{exam-rus} and \ref{exam-rus-machine}). As pointed out in \cite{Pae14}, the sequential quantum circuit in Fig. \ref{fig0} will produce $V_3 \ket{\psi}$ with probability $5/8$ and remain $\ket\psi$ with probability $3/8$, regardless of the exact state $\ket\psi$. Indeed, this assertion can be verified by our algorithm in the sense that given any two different quantum states $\ket{\varphi_1}$ and $\ket{\varphi_2}$, we can conclude that they have the same probability to apply a unitary operator $V_3$ or the identity by verifying $V_\mathit{rus}$-equivalence in $M_\mathit{rus}$, i.e. $\ket{\varphi_1} \sim_{V_\mathit{rus}} \ket{\varphi_2}$, where $V_\mathit{rus} = \text{span} \{ \ket{0}_q\bra{0} \}$. We select $\ket{\varphi_1} = \ket{0}$ and $\ket{\varphi_2} = \ket +$, and $\ket{\varphi_1} = (\ket{0} + i\ket{1})/\sqrt{2}$ and $\ket{\varphi_2} = (\ket{0}+2\ket{1})/\sqrt{5}$ as examples (see Cases qrus-1 and qrus-2 in Table \ref{table1}).

\subsection{Quantum Random Walks}

We consider the quantum random walk example (Example \ref{exam-walk} and \ref{exam-walk-machine}). Our algorithm is applied to the following equivalence checking problems:

{\vskip 3pt}

\textbf{(1) The same coin and different initial states}: Suppose the coin-flip operator $C$ is the Hadamard gate $H$. In this case, the machine is denoted $\mathcal{M}_H$. We consider the two initial states $\ket {0}_c \ket {0}_p$ and $\ket {0}_c \ket {2}_p$. We want to check: whether $\mathcal{M}_H$ with $\ket {0}_c \ket {0}_p$ is equivalent to the same $\mathcal{M}_H$ with $\ket {0}_c \ket {2}_p$? In other words, whether $\ket {0}_c \ket {0}_p \sim \ket {0}_c \ket {2}_p$, i.e. $\ket {0}_c \ket {0}_p \sim_{V_\mathit{walk}} \ket {0}_c \ket {2}_p$ for $V_{\mathit{walk}} = \mathcal{D}(\mathcal{H}_d$)?
To this end, we choose the basis
$
B = \{ \ket {0}_d \bra {0}, \ket {1}_d\bra {1}, \ket {+}_d \bra {+}, \ket {\phi}_d \bra {\phi} \}
$ of $\mathcal{D}(\mathcal{H}_t)$ that consists of only pure states, where
$
\ket \phi = (\ket 0 + i \ket 1)/\sqrt{2}.
$
For convenience, we use $0, 1, \!+\!, \phi$ to denote the corresponding states (i.e., density matrices), and set
$
\rho = \left( \ket {0}_c\bra{0}\otimes \ket {0}_p \bra {0} \right) \oplus \left(- \ket {0}_c\bra{0}\otimes\ket{ 2}_p \bra {2}\right).
$
Now let us see how the algorithm works. It first pushes $(\epsilon, \epsilon)$ into an empty queue $Q$. Repeating the ``while'' loop, it produces set
$
\mathfrak{B} = \left\{ \mathcal{E}_{\epsilon|\epsilon}(\rho), \mathcal{E}_{0|0}(\rho), \mathcal{E}_{1|0}(\rho), \mathcal{E}_{0|\!+\!}(\rho), \dots \right\}.
$
We see that
$\mathcal{E}_{000|\!+\!00}(\rho) = \frac 1 2 \left( \ket {+}_c \bra {+} \otimes \ket {1}_p \bra{1} \right) \oplus 0 \in \mathfrak{B},$
but $\operatorname{tr}(\mathcal{E}_{000|+00}(\rho)) = \frac 1 2 \neq 0$. Therefore, $\ket {0}_c \ket {0}_p$ is not equivalent to $\ket {0}_c \ket {2}_p$ in $\mathcal{M}_H$ (see Case qwalk-small-1 of Table \ref{table1}). In fact,
$
p\left(000 |\!+\!00,\ket {0}_c \ket{0}_p \right) = \frac 1 2 \neq p\left(000 |\!+\!00,\ket {0}_c \ket{2}_p\right) = 0.
$
This example shows an interesting difference between quantum and classical random walks. Note that two initial states $\ket{0}_c\ket{0}_p$ and $\ket{0}_c\ket{2}_p$ are symmetric in the classical case. In the quantum case, however, a superposition of states can be used in a detection. In other words, the Hadamard gate $H$ is biased over $\ket 0$ and $\ket 1$, but we can use $\ket \phi$ as the coin state to overcome this issue \cite{Kem09}. Surprisingly, $\ket {\phi}_c \ket{0}_p$ is equivalent to $\ket {\phi}_c \ket{2}_p$ in $\mathcal{M}_H$, which can also be verified by our algorithm (see Case qwalk-small-2 of Table \ref{table1}).

Furthermore, using our algorithm, we can discover several other interesting symmetries in $\mathcal{M}_H$: $\ket{0}_c\ket{0}_p \sim \ket{1}_c\ket{2}_p$, $\ket{0}_c\ket{1}_p \sim \ket{1}_c\ket{1}_p$ and $\ket{0}_c\ket{2}_p \sim \ket{1}_c\ket{0}_p$ (see Case qwalk-sym-1 to qwalk-sym-3 of Table \ref{table1}).

\begin{remark}
    Our algorithm can be used to test the design of quantum walks against some mathematically  provable properties. For example, since it can be proved  that $\ket{\phi}_c\ket{x}_p \sim \ket{\phi}_c\ket{N-x-2}_p$ for every position $0 \leq x \leq N-2$ in an $N$-dimensional quantum walk with absorbing position $N-1$. Our algorithm can be used to check whether an implementation of quantum walk satisfies this property, and if not, the implementation is incorrect.
\end{remark}

{\vskip 3pt}

\textbf{(2) Two different coins}: The previous example considers different initial states, but the combinational circuits there are the same. Now we replace the Hadamard gate for coin-flip by the gate $
    Y = \frac 1 {\sqrt2} \begin{pmatrix}
        1 & i \\
        i & 1
    \end{pmatrix}$
and the corresponding machine is denoted $\mathcal{M}_Y$; that is, we set $C = Y$ in $\mathcal{M}_C$. It can be verified that $(\mathcal{M}_H, \ket {0}_c \ket{0}_p)$ is equivalent to $(\mathcal{M}_Y, \ket {0}_c \ket{0}_p)$ by our algorithm (see Case qwalk-diff of Table \ref{table1}).

\textbf{(3) Coins with irrational rotations}: Let us further consider two coins with irrational rotations $R_\pm = R_x(\pm\sqrt2)$, where $
    R_x(\theta) = \begin{pmatrix}
        \cos(\theta/2) & -i\sin(\theta/2) \\
        -i\sin(\theta/2) & \cos(\theta/2) \\
    \end{pmatrix}.
$
In this case, the states of the machines may never repeat, as the angles of rotations are irrational. Nevertheless, it can be verified that $(\mathcal{M}_{R_+}, \ket {0}_c \ket{0}_p)$ is equivalent to $(\mathcal{M}_{R_-}, \ket {0}_c \ket{0}_p)$ by our algorithm (see Case qwalk-irr of Table \ref{table1}).

\subsection{Quantum Fourier Transform}

A sequential circuit involving a simple quantum Fourier transform is given in Fig. \ref{fig4}. Suppose there are six qubits $d, q, p_1, p_2, p_3, p_4$, where $d$ is the detective qubit, and $q$ is the control qubit: (1) If the state of $q$ is $\ket 0$, then a $4$-qubit quantum Fourier transform is performed on $p_1, p_2, p_3, p_4$. Recall that an $n$-qubit quantum Fourier transform is defined by
      $
      \mathit{QFT}_n \ket{j} = \frac 1 {\sqrt N} \sum_{k=0}^{N-1} \exp\left( \frac {2\pi ijk} N \right) \ket k$,
      where $N = 2^n$, and for $0 \leq j < N$, $\ket j$ denotes $\ket{j_1 j_2 \dots j_n}$ where $j$ is represented by $\overline{j_1 j_2 \dots j_n}$ in binary form.
  (2) If the state of $q$ is $\ket 1$, then a $\pi/8$ gate $T$ is performed on $p_1$.
\begin{figure}[!htp]\centering
\normalsize\begin{equation*}
\Qcircuit @C=0.7em @R=.5em {
\lstick{d = \ket{0}}        & \qw & \qw & \qw & \qw & \qw   & \qw                        & \qw       & \targ     & \qw & \meter    & \cw  \\
\lstick{q}                  & \qw & \qw & \qw & \qw & \qw   & \ctrlo{1}                  & \ctrl{1}  & \qw       & \qw & \meter    & \cw  \\
\lstick{p_1}                &     & \qw & \qw & \qw & \qw > & \multigate{3}{\text{QFT}}  & \gate{T}  & \qw       & \qw       & \qw & \qw & \qw \\
\lstick{p_2}                & \qwx&     & \qw & \qw & \qw > & \ghost{\text{QFT}}         & \qw       & \qw       & \qw       & \qw & \qw & \qwx \\
\lstick{p_3}                & \qwx& \qwx&     & \qw & \qw > & \ghost{\text{QFT}}         & \qw       & \ctrl{-4} & \qw       & \qw & \qwx & \qwx \\
\lstick{p_4}                & \qwx& \qwx& \qwx&     & \qw > & \ghost{\text{QFT}}         & \qw       & \qw       & \qw       & \qwx& \qwx & \qwx \\
                            & \qwx& \qwx& \qwx& \qwx& \qw   & \qw                        & \qw       & \qw       & \qw \qwx  & \qwx&\qwx & \qwx \\
                            & \qwx& \qwx& \qwx& \qw & \qw   & \qw                        & \qw       & \qw       & \qw       & \qw \qwx&\qwx & \qwx \\
                            & \qwx& \qwx& \qw & \qw & \qw   & \qw                        & \qw       & \qw       & \qw       & \qw & \qw \qwx & \qwx \\
                            & \qwx& \qw & \qw & \qw & \qw   & \qw                        & \qw       & \qw       & \qw       & \qw & \qw & \qw \qwx
                            \gategroup{1}{7}{6}{9}{1.0em}{--}
}
\end{equation*}
\caption{A sequential circuit for the quantum Fourier transform.}\label{fig4}
% \hrulefill \vspace*{4pt}
\end{figure}

At the end of the combinational part, a CNOT gate is performed on $d$ and $p_3$ with $p_3$ being the control qubit so that certain classical information about $p_3$ can be obtained by measuring $d$. Formally, this sequential quantum circuit has $d, q$ as input variables and $p_1, p_2, p_3, p_4$ as state variables, and its combinational part is described by
\begin{equation} \label{eq-qft}
\begin{aligned}
    \mathcal{C}_\mathit{qft}
    =\ & \mathrm{C}^0 \mathit{QFT}_4 [q, p_1, p_2, p_3, p_4]
     \mathrm{C}^1 T [q, p_1] \mathit{CNOT}[p_3, d].
\end{aligned}
\end{equation}

As an illustrative example, we assume the initial state to be either $\ket{0}_{p_1}\ket{0}_{p_2}\ket{0}_{p_3}\ket{0}_{p_4}$ or $\ket{1}_{p_1}\ket{0}_{p_2}\ket{0}_{p_3}\ket{0}_{p_4}$, and we want to determine which is the case. Suppose the input states are only available from a subspace $V_\mathit{qft}$ of $\mathcal{D}(\mathcal{H}_d \otimes \mathcal{H}_q)$:
$
V_\mathit{qft} = \text{span} \{ \ket{0}_d\bra{0} \otimes \ket{0}_q\bra{0}, \ket{0}_d\bra{0} \otimes \ket{1}_q\bra{1} \}.
$
We hope to find an input sequence that can distinguish the two cases if we examine the statistics of output sequences.
Note that at each step there are four possible measurement outcomes, i.e. $0_d0_q$, $0_d1_q$, $1_d0_q$ and $1_d1_q$.
Indeed, our algorithm can find an input sequence $\pi = \ket{0}_d\ket{0}_q, \ket{0}_d\ket{0}_q, \ket{0}_d\ket{1}_q, \ket{0}_d\ket{0}_q$ and present an output sequence $a = 0_d0_q, 0_d0_q, 0_d1_q, 0_d0_q$ such that the two possible initial states behave differently (and therefore we can distinguish them). To be precise, on input sequence $\pi$,  the initial state $\ket{0}_{p_1}\ket{0}_{p_2}\ket{0}_{p_3}\ket{0}_{p_4}$ will produce the  output sequence $a$ with probability $0.338\dots$ while $\ket{1}_{p_1}\ket{0}_{p_2}\ket{0}_{p_3}\ket{0}_{p_4}$ will do so with probability $0.161\dots$ (See Case qft-small-1 of Table \ref{table1}). The gap between the two cases is large enough to distinguish them.

Another example is to distinguish whether the initial state is $\ket{0}_{p_1}\ket{0}_{p_2}\ket{0}_{p_3}\ket{0}_{p_4}$ or $\ket{0}_{p_1}\ket{1}_{p_2}\ket{0}_{p_3}\ket{0}_{p_4}$. Again, our algorithm can find an input sequence $\pi = \ket{0}_d\ket{0}_q,$ $\ket{0}_d\ket{1}_q,$ $\ket{0}_d\ket{0}_q,$ $\ket{0}_d\ket{1}_q,$ $\ket{0}_d\ket{0}_q$ and give an output sequence $a = 0_d0_q, 0_d1_q, 0_d0_q, 0_d1_q, 0_d0_q$ with which the two cases can be distinguished (See Case qft-small-2 of Table \ref{table1}).

\subsection{Quantum Half Adder}

A sequential circuit involving a quantum half adder \cite{Hung04} is given in Fig. \ref{fig5}, where
$
    V = \sqrt X = \frac 1 2 \begin{pmatrix}
        1+i & 1-i \\
        1-i & 1+i
    \end{pmatrix}.
$
There are three qubits $c, a$ and $b$, where $b$ is an accumulator that computes the sum of every $a$ modulo 2, and $c$ contains the carry of $a$ and $b$.
More precisely, this sequential quantum circuit has $c, a$ as input variables and $b$ as the only state variable, and its combinational part is described by
\begin{equation} \label{eq-qha}
\begin{aligned}
    \mathcal{C}_\mathit{qha}
    =\ & \mathrm{C}^1 V^\dag [b, c]
     \mathit{CNOT} [a, b]
     \mathrm{C}^1 V [b, c]
     \mathrm{C}^1 V [a, c] \mathit{CNOT} [a, c].
\end{aligned}
\end{equation}
\begin{figure}[!htp]\centering
\normalsize\begin{equation*}
\Qcircuit @C=0.8em @R=.6em {
\lstick{c}    & \qw & \qw & \qw   & \gate{V^\dag} & \qw       & \gate{V}  & \gate{V}  & \targ & \meter    & \cw  \\
\lstick{a}    & \qw & \qw & \qw   & \qw           & \ctrl{1}  & \qw       & \ctrl{-1} & \ctrl{-1} & \meter & \cw  \\
\lstick{b}    &     & \qw & \qw > & \ctrl{-2}     & \targ     & \ctrl{-2} & \qw       & \qw & \qw \\
              & \qwx& \qw & \qw & \qw         & \qw               & \qw       & \qw       & \qw & \qw \qwx
              \gategroup{1}{5}{3}{9}{1.5em}{--}
}
\end{equation*}
\caption{A sequential quantum circuit for half adders.}\label{fig5}
%\hrulefill \vspace*{4pt}
\end{figure}

A classical (reversible) implementation of this sequential circuit is given in Fig. \ref{fig6}.
The combinational part of sequential classical (reversible) circuit is described by
\begin{equation} \label{eq-ha}
\begin{aligned}
    \mathcal{C}_\mathit{ha}
    = \mathit{Toffoli} [a, b, c] \mathit{CNOT} [a, b].
\end{aligned}
\end{equation}

\begin{figure}[!htp]\centering
\normalsize\begin{equation*}
\Qcircuit @C=.8em @R=.6em {
\lstick{c}    & \qw & \qw & \qw   & \targ         & \qw       & \qw  & \qw & \qw    & \qw & \rstick{c\oplus ab} \\
\lstick{a}    & \qw & \qw & \qw   & \ctrl{-1}     & \ctrl{1}  & \qw  & \qw & \qw & \qw  & \rstick{a} \\
\lstick{b}    &     & \qw & \qw > & \ctrl{-2}     & \targ     & \qw  & \qw & \qw & & \rstick{a\oplus b} \\
              & \qwx& \qw & \qw   & \qw           & \qw              & \qw       & \qw & \qw \qwx
              \gategroup{1}{5}{3}{7}{1.5em}{--}
}
\end{equation*}
\caption{A sequential classical circuit for half adders.}\label{fig6}
% \hrulefill \vspace*{4pt}
\end{figure}

Let us compare the sequential quantum circuit (\ref{eq-qha}) and classical circuit (\ref{eq-ha}) of half adder presented in Fig. \ref{fig5} and Fig. \ref{fig6}. To this end, we should only consider the computational basis, or equivalently work on the subspace:
$ V_{\mathit{qha}} = \text{span}\{\ket{0}_{c}\bra{0}\otimes\ket{0}_{a}\bra{0}, \ket{0}_{c}\bra{0}\otimes\ket{1}_{a}\bra{1}, \ket{1}_{c}\bra{1}\otimes\ket{0}_{a}\bra{0}, \ket{1}_{c}\bra{1}\otimes\ket{1}_{a}\bra{1} \}$.
\iffalse
\begin{align*}
    V_{\mathit{qha}} = \text{span} \{ & \ket{0}_{c}\bra{0}\otimes\ket{0}_{a}\bra{0},
    &\ket{0}_{c}\bra{0}\otimes\ket{1}_{a}\bra{1} &, \\
    &\ket{1}_{c}\bra{1}\otimes\ket{0}_{a}\bra{0},
    &\ket{1}_{c}\bra{1}\otimes\ket{1}_{a}\bra{1} &\}.
\end{align*}
\fi
In checking whether the two sequential (quantum and classical) circuits are (classically functional) equivalent, one would ask, for example, whether $(\mathcal{M}_\mathit{qha}, \ket{0}_b) \sim_{V_{\mathit{qha}}} (\mathcal{M}_\mathit{ha}, \ket{0}_b)$, where $\mathcal{M}_\mathit{qha}$ and $\mathcal{M}_\mathit{ha}$ denote the quantum Mealy machines corresponding to equations (\ref{eq-qha}) and (\ref{eq-ha}), respectively? Indeed, our algorithm returns a positive answer (See Case qhalf-adder of Table \ref{table1}).

\begin{remark} It should be noted that in this case study of half adder, we are comparing a sequential quantum circuit with a classical (reversible) one. In such a special case, the equivalence of them (in terms of sequential quantum circuits)
    %with input states limited to the computational basis
    actually degenerates to that of classical sequential circuits. This is because in this case the input state is always in the computational basis, and the outcome  probability of the sequential quantum circuits must be $0$ or $1$. If a sequential quantum circuit is equivalent to a classical reversible sequential circuit (with input states limited to the computational basis), then they are functionally equivalent. That is, they always behave the same on the same input sequence (in the computational basis).
\end{remark}

\section{Experiments}\label{sec-experiment}

Algorithm \ref{algo} is implemented in Python3 using NumPy package linked with OpenBLAS library. We test our algorithm on a workstation which has an Intel(R) Xeon(R) Platinum 8235 CPU with 16 cores. The
% experimental
results are collected in Table \ref{table1}.

\textbf{An Overview of Test Cases}: Several test cases are based on repeat-until-success scheme (Example \ref{exam-rus}), quantum random walk (Example \ref{exam-walk}), testing circuits for quantum Fourier transform, quantum control circuit and quantum half adder.
%[\textit{The detailed descriptions of the testing circuit for quantum Fourier transform, quantum control circuit and quantum half adder are presented in the Supplemental Material}]

It is difficult to find nontrivial equivalent test cases (that is, two equivalent sequential quantum circuits with different combinational parts). Most of our equivalent test cases are based on the symmetry of quantum random walks, and some other known facts (for example, repeat-until-success scheme).
Non-equivalent test cases are trivial because random sequential quantum circuits are likely to be non-equivalent. However, two random sequential quantum circuits usually reveal their non-equivalence by a very short input sequence. It turns out that it is even difficult to meet two non-equivalent sequential quantum circuits that are equivalent in short input length (formally, they are $K$-equivalent for small $K$). Our nontrivial non-equivalent test cases are based on quantum Fourier transform.
\begin{table}[!htp]
    \centering
    \caption{Experiment results of several test cases.}
\begin{tabular}{|c|c|c|c|c|c|c|}
\hline
Test case     & $d_\mathit{in}$ & $d_{1,2}$ & $d_V$ & $n$ & Eq  & RT(s)           \\ \hline
qwalk-small-1 & 2               & 8         & 4     & 4   & No  & \textless{}0.1 \\ \hline
qwalk-small-2 & 2               & 8         & 4     & 4   & Yes & \textless{}0.1 \\ \hline
qwalk-sym-1   & 2               & 8         & 4     & 4   & Yes & \textless{}0.1 \\ \hline
qwalk-sym-2   & 2               & 8         & 4     & 4   & Yes & \textless{}0.1 \\ \hline
qwalk-sym-3   & 2               & 8         & 4     & 4   & Yes & \textless{}0.1 \\ \hline
qwalk-diff    & 2               & 8         & 4     & 4   & Yes & \textless{}0.1 \\ \hline
qwalk-irr     & 2               & 8         & 4     & 4   & Yes & \textless{}0.1 \\ \hline
qctrl-1       & 8               & 8         & 4     & 6   & Yes & 3.4             \\ \hline
qctrl-2       & 8               & 8         & 4     & 6   & Yes & 3.4             \\ \hline
qhalf-adder   & 4               & 2         & 4     & 3   & Yes & \textless{}0.1 \\ \hline
qft-small-1   & 4               & 16        & 2     & 6   & No  & 0.7             \\ \hline
qft-small-2   & 4               & 16        & 2     & 6   & No  & 1.7             \\ \hline
qrus-1        & 2               & 2         & 1     & 2   & Yes & \textless{}0.1 \\ \hline
qrus-2        & 2               & 2         & 1     & 2   & Yes & \textless{}0.1 \\ \hline
qwalk-large-1 & 2               & 32        & 4     & 6   & Yes & 40.3            \\ \hline
qwalk-large-2 & 2               & 64        & 4     & 7   & Yes & 986.4           \\ \hline
qwalk-large-3 & 2               & 128       & 4     & 8   & Yes & 35117.4             \\ \hline
qft-large-1   & 4               & 32        & 2     & 7   & Yes & 189.6           \\ \hline
qft-large-2   & 4               & 64        & 2     & 8   & No  & 12.5            \\ \hline
qft-large-3   & 4               & 128       & 2     & 9   & No  & 50.9            \\ \hline
\end{tabular}
    \begin{tablenotes}
      \small
      \item $d_\mathit{in}$: the dimension of input Hilbert space.
        $d_{1,2}$: the dimension of state Hilbert space.
        $d_V$: the dimension of available input subspace.
        $n$: the number of qubits.
        Eq: whether $(\mathcal{M}_1, \rho_1) \sim_V (\mathcal{M}_2, \rho_2)$ or not.
        RT(s): the running time of the algorithm.
    \end{tablenotes}
    \label{table1}
\end{table}

\textbf{Efficiency Testing}: Our experiment results on the test cases are summarized in Table \ref{table1}.
We test 6 large test cases, which are qwalk-large-1 to qwalk-large-3 and qft-large-1 to qft-large-3 in Table \ref{table1}. These large test cases are similar to the corresponding small ones, but a larger number of qubits are involved in them.
A big difference between the running times of qwalk-small and qwalk-large occurs due to the exponentially increasing complexity in the number of qubits. For example, the running time of qwalk-large-3 is about 10 hours even if there are only $8$ qubits.
The situation of qft-large is different in that the larger test cases have less running times. This is because the sequential quantum circuits in qft-large-2 and qft-large-3 are not equivalent, and our algorithm is designed (for efficiency) to halt as soon as it finds a witness that shows the non-equivalence of the circuits.

\section{Conclusion}\label{sec-con}

This paper defines a Mealy machine-based framework and develops an algorithm for equivalence checking of sequential quantum circuits. To the best of our knowledge, it is the first algorithm for this purpose. The complexity of this algorithms is $\mathcal{O}(2^{3m+5l}(2^{3m}+2^{3l}))$ when applied to two circuits with $m$ input qubit variables and $l_1, l_2$ state qubit variables, respectively.
Although the gap between it and the complexity of the best known algorithm for checking equivalence of classical sequential circuits is not very large (see Table \ref{table2}), there is still some possibility to improve our algorithm.
So, an important issue for future research is to find more efficient algorithms for equivalence checking of sequential quantum circuits.

It is widely believed that in the near future, only NISQ (Noisy Intermediate-Scale Quantum) devices are physically realisable and they will find real-world applications in such areas as quantum chemistry and many-body physics  \cite{Pre18}. So, another interesting research topic is to develop algorithms for checking approximate equivalence between sequential quantum circuit and their noisy implementations. But the nature of this problem seems very different from the exact equivalence checking studied in this paper, as hinted by the results in \cite{Tze92} for probabilistic automata.

Model checking has been successfully applied to verify some sophisticated properties of very large classical sequential circuits; see for example \cite{Bur94}.
In the last fifteen years, model checking techniques have been extended for verification of quantum communication protocols \cite{Gay08} \cite{Dav12} \cite{Gay18}, quantum automata \cite{YL14} and quantum Markov chains \cite{YSG}.
So, another interesting topic is to see how these quantum model checking techniques can be tuned
for checking sequential quantum circuits.

Up to now, all of the algorithms for checking equivalence of (both combinational and sequential) quantum circuits are classical (i.e. to be run on classical computers). An even more interesting topic is to find \textit{quantum} algorithms for equivalence checking of quantum circuits so that the existing small quantum computers can be used in designing larger and larger ones.

\section*{Acknowledgment}

We thank Prof. Yuan Feng for pointing out a mistake in an early version of this paper.

\addcontentsline{toc}{section}{References}

\bibliographystyle{unsrt} \bibliography{arxiv}

\appendix

\section{Hardware Realisation of Sequential Quantum Circuits} \label{sec:feasibility}
To show that the sequential quantum circuit model is physically feasible and meaningful, we briefly discuss its realization on real quantum hardware in this section. We will take superconducting quantum computing as an example, which is now one of the most advanced and promising technique paths for quantum computers.

Different from classical circuits, quantum gates in superconducting quantum computers are realized by specific pulse signals instead of concrete electronic components, so quantum data will not ``flow through'' the quantum gates, and conversely, the gate pulses are applied on the quantum data, which are stored in quantum registers (namely, qubits). Quantum hardware itself needs a clock to ensure correct timing for gate pulses, otherwise the computing results will be totally unpredictable. Despite these differences, sequential quantum circuits are still realizable on superconducting quantum hardware.

The realization of sequential quantum circuits requires dynamic quantum computing \cite{Rya17}. Dynamic quantum computing allows feedforward and feedback, which means that qubits could be measured intermediately during the computation and the measurement outcomes might in return affect the following computing. A common approach to realize dynamic quantum computing is to design a control system for quantum processors using field-programmable gate array (FPGA) \cite{Rol19}\cite{Rol20}.

\begin{figure}[!hbp]
    \centering
    \includegraphics[scale=0.25]{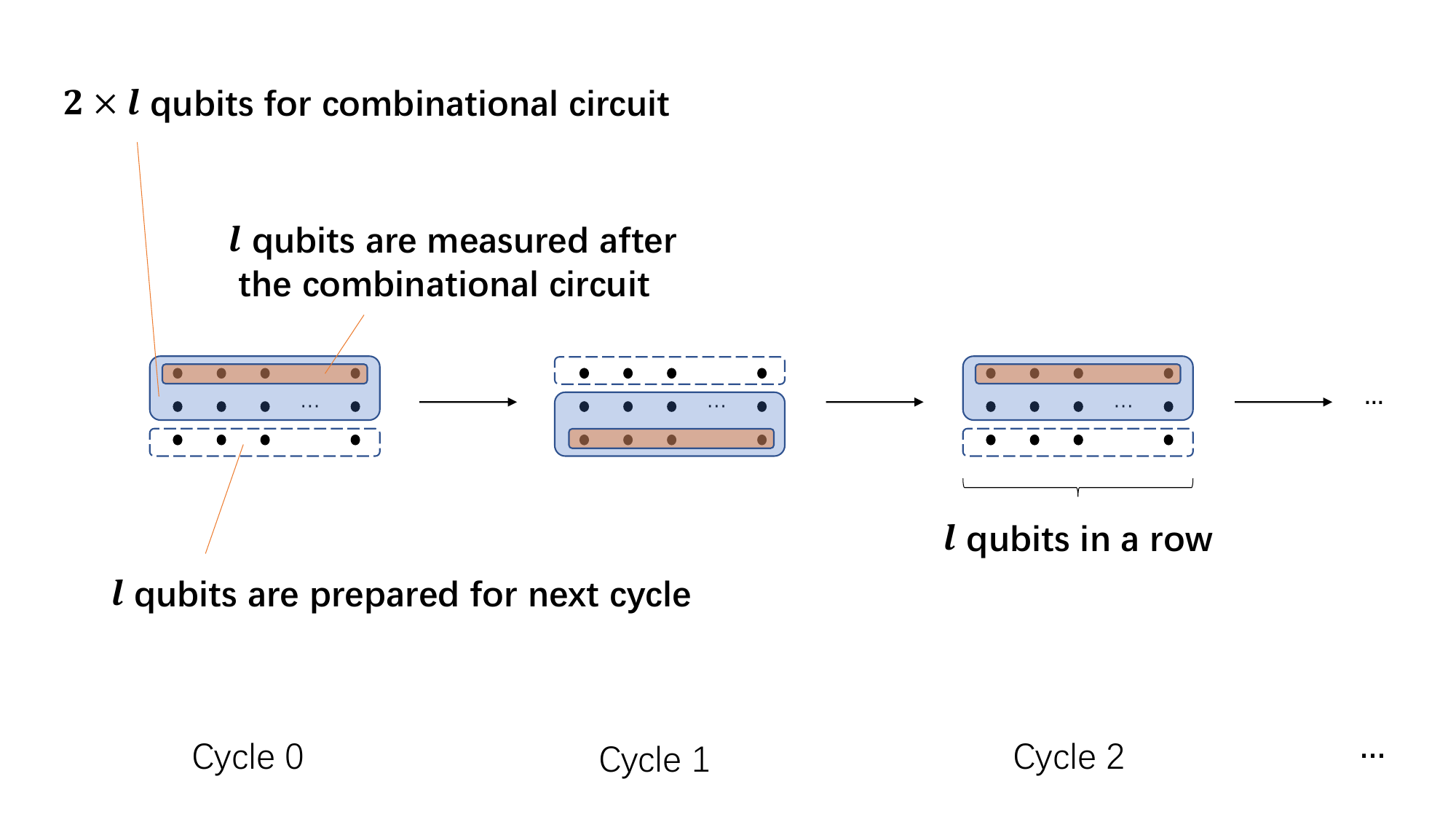}
    \caption{A sequential circuit implemented on a 2D-grid superconducting quantum processor with $3l$ qubits. Here we set $m=l$. The $l$ qubits in middle row represent the $l$ state variables, while the $l$ qubits in the top row and $l$ qubits in the bottom row alternatively represent the $m$ ($m = l$) input variables. At Cycle $0$, the first $l$ qubits are input variables and they together with the second row are applied by a combinational circuit, and after that their measurement outcomes are passed to the $l$ output variables. The bottom row in the dashed frame is prepared as input for Cycle 1. At Cycle 1, the qubits in the bottom row become the input variables, while the qubits in the top row are reset to prepare the input for Cycle $2$. During the whole computing process, the $l$ qubits in middle row are always active.}
    \label{fig:seqC_on_superConducting}
\end{figure}

In our sequential quantum circuit model, it is only required that the rest qubits are still working after some qubits are measured, rather than a complete feedforward and feedback control mechanism.
As stated in Section \ref{defSQC}, a synchronous clock is needed in a sequential quantum circuit to synchronize the computation, which is different from the internal clock of FPGA control system (whose frequency is usually 100 MHz or higher) to guarantee correct gate pulses. A single-qubit or two-qubit unitary gate usually costs about 10 to 40 nanoseconds \cite{Chen14} \cite{Aru19}. Measurement operations usually take longer (about a few hundred nanoseconds) \cite{Rya17}. What we need is to design a synchronous clock with one-cycle duration longer than the whole time consumption of both the combination and measurement parts in the sequential quantum circuit (besides, the one-cycle duration of this clock should be a multiple of the one-cycle duration of the internal clock). Apparently this kind of synchronous clock is feasible and can be added to the control system. Fig. \ref{fig:seqC_on_superConducting} shows an example of sequential quantum circuits with $n=2l$ qubits executed on an superconducting quantum processor with $3l$ qubits.

\section{Completing the Proof of Theorem \ref{thm-eq}} \label{app:thm-eq}

Here, we provide the details of steps (i) and (ii) outlined at the beginning of the proof of Theorem \ref{thm-eq}.
These two steps will be carried out through a series of lemmas. We start from giving a technical lemma that gives explicit formulas for computing the transition probability and the machine's state at each step.

\begin{lemma} \label{lem-prob}
    Let $\mathcal{M} = \{ \mathcal{H}_\mathit{in}, \mathcal{H}_s, U, M \}$ be a quantum Mealy machine. Let $\mathcal{E}_{a|\pi}$ be the super-operator defined by equation (\ref{eq-superoperator}) and generalised by equation (\ref{eq-composition}). Then for every initial state $\rho_0 \in \mathcal{D}(\mathcal{H}_s)$, input sequence $\pi$ in $\mathcal{D}(\mathcal{H}_\mathit{in})$ and output sequence $a \in O^{\abs{\pi}}$, we have
    \[
        p(a|\pi, \rho_0) = \operatorname{tr} \left[ \mathcal{E}_{a|\pi}(\rho_0) \right],
    \]
    and the density operator on input sequence $\pi$ and output sequence $a$ is
    \[
        \rho_{\abs{\pi}} = \frac 1 {\operatorname{tr} \left[ \mathcal{E}_{a|\pi}(\rho_0) \right]} \mathcal{E}_{a|\pi}(\rho_0).
    \]
\end{lemma}

\textit{Proof}:
    We prove it by induction on the length of $\pi$. It is trivial for the case that $\pi$ is empty, i.e. $\abs{\pi} = 0$. Suppose it is true for the case $\abs{\pi} = k$, i.e.
    \[
        p(a|\pi, \rho_0) = \operatorname{tr} \left[ \mathcal{E}_{a|\pi}(\rho_0) \right]
    \]
    for every initial state $\rho_0 \in \mathcal{D}(\mathcal{H}_s)$, input sequence $\pi$ in $\mathcal{D}(\mathcal{H}_\mathit{in})$ with $\abs{\pi} = k$ and output sequence $a \in O^{\abs{\pi}}$. The density operator on input sequence $\pi$ and output sequence $a$ is
    \[
        \rho_k = \frac 1 {\operatorname{tr} \left[ \mathcal{E}_{a|\pi}(\rho_0) \right]} \mathcal{E}_{a|\pi}(\rho_0).
    \]
    Let the next input and output be $\sigma_{k+1}$ and $a_{k+1}$, respectively. Then the probability with output $a_{k+1}$ on input $\sigma_{k+1}$ is
    \begin{align*}
        p(a_{k+1}|\sigma_{k+1}, \rho_k)
        & = \operatorname{tr} \left[ \mathcal{E}_{a_{k+1}| \sigma_{k+1}}(\rho) \right] \\
        & = \frac 1 {\operatorname{tr} \left[ \mathcal{E}_{a|\pi}(\rho_0) \right]} \operatorname{tr}\left[\mathcal{E}_{aa_{k+1}|\pi\sigma_{k+1}}(\rho_0)\right].
    \end{align*}
    By the definition, the probability on input sequence $\pi \sigma_{k+1}$ and output sequence $aa_{k+1}$ is
    \begin{align*}
        p(aa_{k+1}|\pi\sigma_{k+1}, \rho_0)
        & = p(a|\pi, \rho_0) p(a_{k+1}|\sigma_{k+1}, \rho_k) \\
        & = \operatorname{tr}\left[\mathcal{E}_{aa_{k+1}|\pi\sigma_{k+1}}(\rho_0)\right],
    \end{align*}
    and the density operator is
    \begin{align*}
        \rho'
        & = \frac 1 {p(a_{k+1}|\sigma_{k+1}, \rho_k)} \mathcal{E}_{a_{k+1}| \sigma_{k+1}}(\rho) \\
        & = \frac 1 {\operatorname{tr}\left[\mathcal{E}_{aa_{k+1}|\pi\sigma_{k+1}}(\rho_0)\right]} \mathcal{E}_{aa_{k+1}|\pi\sigma_{k+1}}(\rho_0).
    \end{align*}
    Hence, the property also holds for the case $\abs{\pi} = k+1$, and the proof is completed.
\hfill $\blacksquare$

The next lemma presents a characterisation of the dynamics of the direct sum of two machines in terms of their respective dynamics.

\begin{lemma} \label{lem-superoperator}
    Let $\mathcal{M}_i = \{ \mathcal{H}_\mathit{in}, \mathcal{H}_s^{(i)}, U_i, M_i \}$ ($i = 1, 2$) be two quantum Mealy machines with the same input space $\mathcal{H}_\mathit{in}$ and the same outputs $O$ (i.e. measurements $M_i = \left\{ M_a^{(i)}: a \in O \right\}$ for $i = 1, 2$). Let $\mathcal{E}_{a|\pi}^{(1)}, \mathcal{E}_{a|\pi}^{(2)}$ and $\mathcal{E}_{a|\pi}$ be the super-operator defined in $\mathcal{M}_1, \mathcal{M}_2$ and $\mathcal{M} = \mathcal{M}_1 \oplus \mathcal{M}_2$ by equation (\ref{eq-superoperator}) and generalised by equation (\ref{eq-composition}), respectively. Then
    \[
        \mathcal{E}_{a|\pi}(\rho_1 \oplus \rho_2) = \mathcal{E}_{a|\pi}^{(1)}(\rho_1) \oplus \mathcal{E}_{a|\pi}^{(2)}(\rho_2)
    \]
    for every $\rho_i \in \mathcal{D}(\mathcal{H}_s^{(i)})$ ($i = 1, 2$).
\end{lemma}

\textit{Proof}: We prove this lemma by induction on the length of $\pi$. For the case of $\abs{\pi} = 1$, for $a \in O$ and $\sigma \in \mathcal{D}(\mathcal{H}_\mathit{in})$, we can verify that
\begin{align*}
    \mathcal{E}_{a|\sigma}(\rho_1 \oplus \rho_2)
    & = \operatorname{tr}_{\mathcal{H}_\mathit{in}} \Big[ \left(M_a^{(1)} \oplus M_a^{(2)}\right) \left( U_1 \oplus U_2 \right) \\
    & \quad \left( \sigma \otimes \left( \rho_1 \oplus \rho_2 \right) \right) \left( U_1 \oplus U_2 \right)^\dag
     \left(M_a^{(1)} \oplus M_a^{(2)}\right) ^\dag \Big] \\
    & = \mathcal{E}_{a|\sigma}^{(1)}(\rho_1) \oplus \mathcal{E}_{a|\sigma}^{(2)}(\rho_2).
\end{align*}
Now suppose it is true for the case of $\abs{\pi} = k$, i.e.
\[
    \mathcal{E}_{a|\pi}(\rho_1 \oplus \rho_2) = \mathcal{E}_{a|\pi}^{(1)}(\rho_1) \oplus \mathcal{E}_{a|\pi}^{(2)}(\rho_2)
\]
holds for every input sequence $\pi$ with $\abs{\pi} = k$ and output sequence $a$. For every $a_{k+1} \in O$ and $\sigma_{k+1} \in \mathcal{D}(\mathcal{H}_\mathit{in})$, we have
\begin{align*}
    \mathcal{E}_{aa_{k+1}|\pi\sigma_{k+1}}(\rho_1 \oplus \rho_2)
    & = \mathcal{E}_{a_{k+1}|\sigma_{k+1}} \left( \mathcal{E}_{a|\pi}(\rho_1 \oplus \rho_2) \right) \\
    & = \mathcal{E}_{a_{k+1}|\sigma_{k+1}} \left( \mathcal{E}_{a|\pi}^{(1)}(\rho_1) \oplus \mathcal{E}_{a|\pi}^{(2)}(\rho_2) \right) \\
    & = \mathcal{E}_{aa_{k+1}|\pi\sigma_{k+1}}^{(1)}(\rho_1) \oplus \mathcal{E}_{aa_{k+1}|\pi\sigma_{k+1}}^{(2)}(\rho_2).
\end{align*}
Hence, the property also holds for the case $\abs{\pi} = k+1$, and we complete the proof.
\hfill $\blacksquare$

Now we are ready to accomplish step (i). Formally, this step can be formulated as the following:

\begin{lemma} \label{lem-directsum}
    Let $\mathcal{M}_i = \{ \mathcal{H}_\mathit{in}, \mathcal{H}_s^{(i)}, U_i, M_i \}$ ($i = 1, 2$) be two quantum Mealy machines with the same input space $\mathcal{H}_\mathit{in}$ and the same outputs $O$ (i.e. measurements $M_i = \left\{ M_a^{(i)}: a \in O \right\}$ for $i = 1, 2$). Let $\rho_i \in \mathcal{D}(\mathcal{H}_s^{(i)})$ ($i = 1, 2$) and $V \subseteq \mathcal{D}(\mathcal{H}_\mathit{in})$. Then $(\mathcal{M}_1, \rho_1) \sim_V (\mathcal{M}_2, \rho_2)$ iff $\rho_1 \sim_V \rho_2$ in $\mathcal{M}_1 \oplus \mathcal{M}_2$.
\end{lemma}
\textit{Proof}:
    Let $\mathcal{M} = \mathcal{M}_1 \oplus \mathcal{M}_2$. Furthermore,
    let $\mathcal{E}_{a|\pi}^{(i)}$ be the super-operator defined in $\mathcal{M}_i$ for $i = 1, 2$, and $\mathcal{E}_{a|\pi}$ be that defined in $\mathcal{M}$. We also use the notation $p^{(i)}(a|\pi, \rho) (i=1,2)$ and $p(a|\pi, \rho)$ to denote the probability that $\mathcal{M}_i(i = 1, 2)$ and $\mathcal{M}$ with initial state $\rho$ output $a$ on input sequence $\pi$, respectively. Then by Lemma \ref{lem-prob}, we have:
    \[
        p^{(i)}(a|\pi, \rho_i) = \operatorname{tr}\left[ \mathcal{E}_{a|\pi}^{(i)}(\rho_i) \right],\qquad
        p(a|\pi, \rho_i) = \operatorname{tr}\left[ \mathcal{E}_{a|\pi}(\rho_i) \right]
    \]
    for $i = 1, 2$. By Lemma \ref{lem-superoperator}, we have:
    $
        \mathcal{E}_{a|\pi}(\rho_i) = \mathcal{E}_{a|\pi}^{(i)}(\rho_i)
    $
    for $i = 1, 2$. We immediately obtain that
    $
        p(a|\pi, \rho_i) = p^{(i)}(a|\pi, \rho_i),
    $
    which implies that $(\mathcal{M}_1, \rho_1) \sim_V (\mathcal{M}_2, \rho_2)$ iff $\rho_1 \sim_V \rho_2$ in $\mathcal{M}_1 \oplus \mathcal{M}_2$ by the definition of $V$-equivalence.
\hfill $\blacksquare$

Next, we turn to accomplish step (ii). Formally, we have:
\begin{lemma} \label{lem-eqbasis}
    Let $\mathcal{M} = (\mathcal{H}_\mathit{in}, \mathcal{H}_s, U, M)$ be a quantum Mealy machine, and $B \subseteq \mathcal{D}(\mathcal{H}_\mathit{in})$.
    Then $\rho_1\sim_B \rho_2$ iff $\rho_1\sim_{\operatorname{span} B}\rho_2$ for every $\rho_1, \rho_2 \in \mathcal{D}(\mathcal{H}_s)$.
\end{lemma}

\textit{Proof}: The \textquotedblleft if\textquotedblright\ part is obvious. To prove the  \textquotedblleft only if\textquotedblright\ part, we first observe that
    for every input sequence $\pi$ and $a \in O^{|\pi|}$,
    $
    p(a|\pi, \rho) = \operatorname{tr}\left[\mathcal{E}_{a|\pi}(\rho)\right].
    $
    We only need to consider the case where $B$ is a finite set, say $B = \{ \gamma_1, \gamma_2, \dots, \gamma_{k} \}$. Suppose that $\rho_1\sim_B \rho_2$. Then for every input sequence $\pi = \sigma_1 \sigma_2 \dots \sigma_t$ and $a \in O^t$, where $\sigma_j \in \operatorname{span} B$ for $1 \leq j \leq t$, we have
    \[
    \sigma_j = \sum_{i=1}^{k} \alpha_{ji} \gamma_i.
    \]
    for some coefficients $\alpha_{ji}$. By Lemma \ref{lem-prob}, we have
    \begin{align*}
        p(a|\pi, \rho_1)
        & = \operatorname{tr}\left[\mathcal{E}_{a|\pi}(\rho_1)\right] \\
        & = \sum_{i_1 = 1}^{k} \dots \sum_{i_t = 1}^{k} \alpha_{1i_1} \dots \alpha_{ti_t} \operatorname{tr}\left[\mathcal{E}_{a|\gamma_{i_1}\dots\gamma_{i_t}}(\rho_1)\right] \\
        & = \sum_{i_1 = 1}^{k} \dots \sum_{i_t = 1}^{k} \alpha_{1i_1} \dots \alpha_{ti_t} \operatorname{tr}\left[\mathcal{E}_{a|\gamma_{i_1}\dots\gamma_{i_t}}(\rho_2)\right] \\
        & = \operatorname{tr}\left[\mathcal{E}_{a|\pi}(\rho_2)\right]
         = p(a|\pi, \rho_2),
    \end{align*}
    which yields $\rho_1\sim_{\operatorname{span} B}\rho_2$. \hfill $\blacksquare$

As an immeduate corollary, we see that $\rho_1 \sim \rho_2$ if and only if
     $\rho_1 \sim_B \rho_2$ for some basis $B$ of $\mathcal{D}(\mathcal{H}_\mathit{in})$ consisting of only pure states.

\section{Correctness Proof of Algorithm \ref{algo}} \label{app:correctness}

The correctness of Algorithm 1 is essentially based on Theorem \ref{thm-eq}. Indeed, by Theorem \ref{thm-eq} we only need to prove that $\operatorname{span} D(\rho, d_1^2+d_2^2-1) = \operatorname{span} \mathfrak{B}.$ The inclusion $\operatorname{span} D(\rho, d_1^2+d_2^2-1) \supseteq \operatorname{span} \mathfrak{B}$ is obvious. To prove the reverse inclusion, it is sufficient to show that $\mathcal{E}_{a|\pi}(\rho) \in \operatorname{span} \mathfrak{B}_{\leq (\pi, a)}$ for every $\mathcal{E}_{a|\pi}(\rho) \in D(\rho, d_1^2+d_2^2-1),$ where $$\mathfrak{B}_{\lhd(\pi, a)} = \{ \mathcal{E}_{a'|\pi'}(\rho): (\pi', a') \lhd (\pi, a) \text{ and } \mathcal{E}_{a'|\pi'}(\rho) \in \mathfrak{B} \}$$ and $\lhd$ can be $<$ or $\leq$. This can be proved by induction on the admissible order maintained in the queue. Note that $\mathcal{E}_{\epsilon|\epsilon}(\rho) = \rho \in \operatorname{span} \mathfrak{B}$ for the basis. For every $\mathcal{E}_{ax|\pi\sigma}(\rho)$, we consider the following two cases:

{\vskip 4pt}

    \textbf{Case 1}. If $\mathcal{E}_{a|\pi}(\rho)$ is added into $\mathfrak{B}$ at the step when $(\pi, a)$ is popped from $Q$ in the algorithm, then $\mathcal{E}_{ax|\pi\sigma}(\rho)$ is pushed into $Q$ in the execution of the algorithm and thus $\mathcal{E}_{ax|\pi\sigma}(\rho) \in \operatorname{span} \mathfrak{B}_{\leq (\pi\sigma, ax)}$ is guaranteed.

    {\vskip 4pt}

    \textbf{Case 2}. If $\mathcal{E}_{a|\pi}(\rho)$ is not added into $\mathfrak{B}$ at the step when $(\pi, a)$ is popped from $Q$ in the algorithm or even $(\pi, a)$ is never in the queue, then $\mathfrak{B}_{< (\pi, a)} = \mathfrak{B}_{\leq (\pi, a)}.$ By the induction hypothesis we have: $\mathcal{E}_{a|\pi}(\rho) \in \operatorname{span} \mathfrak{B}_{<(\pi, a)},$ which immediately yields
    \begin{align*}
        \mathcal{E}_{ax|\pi\sigma}(\rho)
        & = \mathcal{E}_{x|\sigma} ( \mathcal{E}_{a|\pi} (\rho) ) \\
        & \in \mathcal{E}_{x|\sigma} ( \operatorname{span} \mathfrak{B}_{<(\pi, a)} ) \\
        & = \operatorname{span} \mathcal{E}_{x|\sigma} ( \mathfrak{B}_{<(\pi, a)} ) \\
        & \subseteq \operatorname{span} \mathfrak{B}_{<(\pi\sigma, ax)} \\
        & \subseteq \operatorname{span} \mathfrak{B}_{\leq (\pi\sigma, ax)}.
    \end{align*}
    Hence, $\mathcal{E}_{ax|\pi\sigma}(\rho) \in \operatorname{span} \mathfrak{B}_{\leq (\pi\sigma, ax)}$ is also guaranteed.

    {\vskip 4pt}

    Therefore, it always holds that $\operatorname{span} D(\rho, d_1^2+d_2^2-1) \subseteq \operatorname{span} \mathfrak{B}$ and correctness of the algorithm is proved.

\end{document}